\documentclass[journal=jacsat,manuscript=article]{achemso}
\usepackage[version=4]{mhchem} 
\usepackage[T1]{fontenc}       
\setkeys{acs}{articletitle = true}
\usepackage[labelfont=bf,font=small,labelsep=period,justification=raggedright]{caption}

\usepackage{changepage}

\usepackage{amsmath,amssymb}

\usepackage{mathtools}

\usepackage[utf8]{inputenc}

\usepackage{units, xfrac}

\allowdisplaybreaks

\SectionNumbersOn 

\usepackage{nameref}
\usepackage{hyperref}

\usepackage{enumitem}
\usepackage{graphicx}
\usepackage{amssymb}

\usepackage[euler]{textgreek}

\usepackage{color}
\usepackage[table,xcdraw,dvipsnames]{xcolor}

\usepackage{setspace}

\author{Vahe Galstyan}
\altaffiliation{Contributed equally to this work}
\affiliation{Biochemistry and Molecular Biophysics Option, California Institute of Technology, Pasadena, California 91125, United States}
\author{Luke Funk}
\altaffiliation{Contributed equally to this work}
\affiliation{Harvard-MIT Division of Health Sciences and Technology and the Broad Institute of MIT and Harvard, Massachusetts Institute of Technology, Cambridge, Massachusetts 02139, United States}
\author{Tal Einav}
\affiliation{Department of Physics, California Institute of Technology, Pasadena, California 91125, United States}
\author{Rob Phillips}
\affiliation{Department of Physics, California Institute of Technology, Pasadena, California 91125, United States}
\alsoaffiliation{Department of Applied Physics, California Institute of Technology, Pasadena, California 91125, United States}
\alsoaffiliation{Division of Biology and Biological Engineering, California Institute of Technology, Pasadena, California 91125, United States}
\email{phillips@pboc.caltech.edu}
\phone{(626) 395-3374}

\title{Combinatorial Control through Allostery}

\begin{document}

\maketitle
\singlespacing

\addtocontents{toc}{\protect\setcounter{tocdepth}{0}}

\pagebreak
\begin{abstract}
	Many instances of cellular signaling and transcriptional regulation involve
	switch-like molecular responses to the presence or absence of input ligands.
	To understand how these responses come about and how they can be harnessed,
	we develop a statistical mechanical model to characterize the types of Boolean
	logic that can arise from allosteric molecules following the Monod-Wyman-Changeux
	(MWC) model. Building upon previous work, we show how an allosteric
	molecule regulated by two inputs can elicit AND, OR, NAND and NOR responses,
	but is unable to realize XOR or XNOR gates.
	Next, we demonstrate the ability of an MWC molecule to perform ratiometric sensing
	- a response behavior where activity depends monotonically on the ratio of ligand
	concentrations.
	We then extend our analysis to more general schemes of combinatorial control involving
	either additional binding sites for the two ligands or an additional third ligand and show
	how these additions can cause a switch in the logic behavior of the molecule.
	Overall, our results demonstrate the wide variety of control schemes that biological systems can
	implement using simple mechanisms.
\end{abstract}

\section*{Introduction}

A hallmark of cellular signaling and regulation is combinatorial control. 
Disparate examples ranging from metabolic enzymes to actin polymerization to
transcriptional regulation involve multiple inputs that
often give rise to a much richer response than what could be achieved through a
single-input. For example, the bacterial enzyme
phosphofructokinase in the glycolysis pathway is allosterically regulated by both ADP and PEP \cite{Blangy1968}. Whereas PEP serves as an allosteric inhibitor, ADP is both an allosteric activator and a competitive inhibitor
depending upon its concentration. 
This modulation by multiple allosteric ligands gives rise to a complex control of the flux through the glycolytic pathway: 
increasing ADP concentration first increases the activity of phosphofructokinase (via the
allosteric modulation) but ultimately decreases it (from competitive inhibition).
Another example is offered by the polymerization of actin  at the leading edge of  motile cells.
In particular, the presence of two ligands, Cdc42 and PIP2, 
is required to activate the protein N-WASP by binding to it in a way that  
permits it to then activate the Arp2/3 complex and stimulate actin polymerization.
\cite{DueberLimScience2003}

\begin{figure}[!ht]
	\centerline{\includegraphics{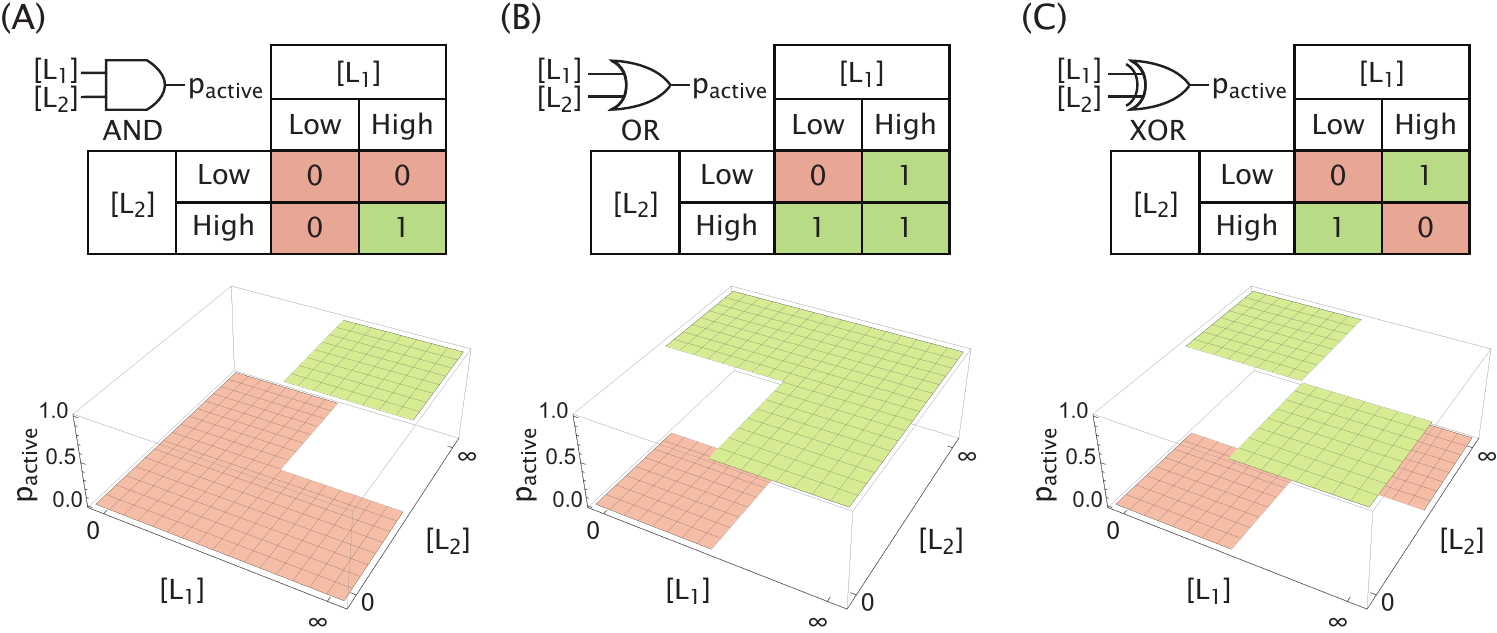}}
	\caption{\textbf{Logic gates as molecular responses.} The (A) AND, (B) OR, and
		(C) XOR gates are represented through their corresponding logic tables as well
		as target activity profiles regulated by two ligands.
		The behavior of each gate is measured solely by its activity in the absence and
		at saturating concentrations of each ligand and not by the character of the
		active/inactive transition.} \label{fig:logic_gates}
\end{figure}

In the context of transcriptional regulation, an elegant earlier work explored
the conditions under which transcriptional regulatory networks could give rise
to the familiar Boolean logic operations, like those shown in
Figure~\ref{fig:logic_gates} \cite{Buchler2003}. There it was found that the
combined effect of two distinct transcription factors on the transcriptional
activity of a given promoter depend upon their respective binding strengths as
well as the cooperative interactions between each other and the RNA polymerase.
Indeed, by tuning the binding strengths and cooperativity parameters, one could
generate a panoply of different logic gates such as the familiar AND, OR, NAND
(NOT-AND) and NOR (NOT-OR) gates, known from the world of digital electronics. \cite{Buchler2003}

Here we explore the diversity of combinatorial responses 
that can be effected by a single allosteric molecule
by asking if such molecules can yield multi-input combinatorial control
in the same way that transcriptional networks have already been shown to.
Specifically, we build on earlier work that shows that an allosteric molecule described by the
Monod-Wyman-Changeux (MWC) model can deliver input-output functions similar to
the ideal logic gates described in Figure~\ref{fig:logic_gates}. \cite{Graham2005, deRonde2012, Agliari2015}
In the MWC model, an allosteric molecule 
exists in a thermodynamic equilibrium between active and inactive 
states, with the relative occupancy of each state being modulated 
by regulatory ligands. \cite{MartinsSwainPLoSCompBiol2011}
We use statistical mechanics to characterize the input-output response of such a
molecule in the limits where each of the two ligands is either absent or at a saturating
concentration and determine the necessary conditions to form the various logic
gates, with our original contribution on this point focusing on a systematic 
exploration of the MWC parameter space for each logic gate.

We then analyze the MWC response modulated by two input ligands but
outside of traditional Boolean logic functions. In particular, 
we show how, by tuning the MWC parameters, 
the response (probability of the allosteric protein being active)
 in any three of the four concentration limits 
can be explicitly controlled,
along with the ligand concentrations at which transitions between
these limit responses occur.
Focusing next on the profile of the response near the transition
concentrations, we demonstrate how an MWC molecule 
can exhibit ratiometric sensing which was 
observed experimentally in the bone morphogenetic protein (BMP) 
signaling pathway\cite{Antebi2017} as well as in galactose 
metabolic (GAL) gene induction in yeast\cite{EscalanteChong2015}.

Additionally, we extend our analysis of logic responses to cases
beyond two-ligand control with a single binding site for each ligand.
We first discuss the effect of the number of binding sites on the logic response
and demonstrate how altering that number, which can occur through evolution or
synthetic design, is able to cause a switch in the logic-behavior of
an MWC molecule, such as transitioning from AND into OR behavior.
Next, we explore the increased diversity of logic responses that can be
achieved by three-ligand MWC molecules compared with the two-ligand case
and offer an interesting
perspective on the role of the third ligand as a regulator that can
switch the logic-behavior formed by the other two ligands.
We end by a discussion of our theoretical results in the context of a
growing body of experimental works on 
natural and \textit{de novo} designed molecular logic gates.
In total, these results hint at simple mechanisms that biological
systems can utilize to refine their combinatorial control.

\section*{Results}
\subsection*{Logic Response of an Allosteric Protein Modulated by Two Ligands}

\begin{figure}[t]
	\centerline{\includegraphics{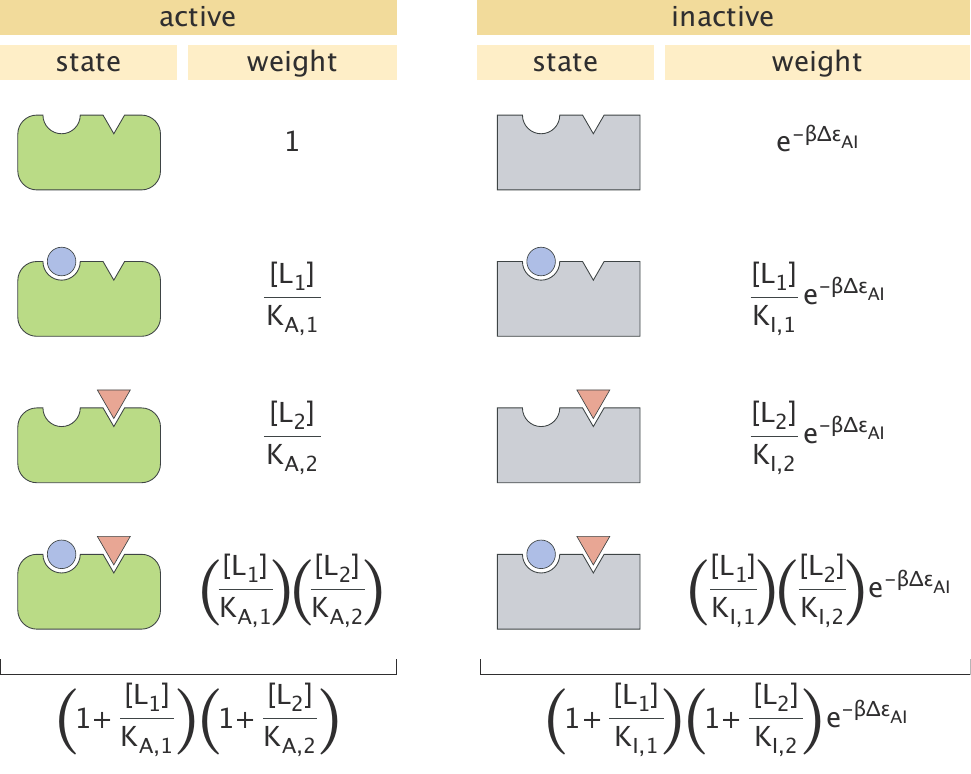}}
	\caption{\textbf{States and weights for the allosteric protein.} The two
	different ligands (blue circle ($\text{i}=1$) and red triangle ($\text{i}=2$)) are present at concentrations
	$[\text{L}_\text{i}]$ and with a dissociation constant
	$\text{K}_{\text{A},\text{i}}$ in the active state and
	$\text{K}_{\text{I},\text{i}}$ in the inactive state. The
	energetic difference between the inactive and active states is
	denoted by $\Delta \text{\textepsilon}_{\text{AI}} = \text{\textepsilon}_{\text{I}} -
	\text{\textepsilon}_{\text{A}}$. Total weights of the active and inactive states are
	shown below each column and are obtained by summing all the weights in
	that column.} \label{fig:states_weights}
\end{figure}

Consider an MWC molecule, as shown in Figure~\ref{fig:states_weights}, that
fluctuates between active and inactive states (with $\Delta
\text{\textepsilon}_{\text{AI}}$ defined as the free energy difference between the inactive
and active states in the absence of ligand). We enumerate the entire set of
allowed states of activity and ligand occupancy, along with their corresponding
statistical weights. The probability that this protein is active depends on the
concentrations of two input molecules, $[\text{L}_1]$ and $[\text{L}_2]$, and is
given by
\begin{equation} \label{eq:mwc}
\text{p}_{\text{active}}\left( [\text{L}_1], [\text{L}_2] \right) = \frac{\left( 1 + \frac{[\text{L}_1]}{\text{K}_{\text{A},1}} \right)\left( 1 + \frac{[\text{L}_2]}{\text{K}_{\text{A},2}} \right)}{\left( 1 + \frac{[\text{L}_1]}{\text{K}_{\text{A},1}} \right)\left( 1 + \frac{[\text{L}_2]}{\text{K}_{\text{A},2}} \right) + \text{e}^{-\text{\textbeta} \Delta \text{\textepsilon}_{\text{AI}}} \left( 1 + \frac{[\text{L}_1]}{\text{K}_{\text{I},1}} \right)\left( 1 + \frac{[\text{L}_2]}{\text{K}_{\text{I},2}} \right)},
\end{equation}
where $\text{K}_{\text{A},\text{i}}$ and $\text{K}_{\text{I},\text{i}}$ are the
dissociation constants between the $\text{i}^\text{th}$ ligand and the active or
inactive protein, respectively.  We begin with the two-input case
such that  $\text{i}=1 \text{ or } 2$.

\begin{figure}[!ht]
	\centerline{\includegraphics{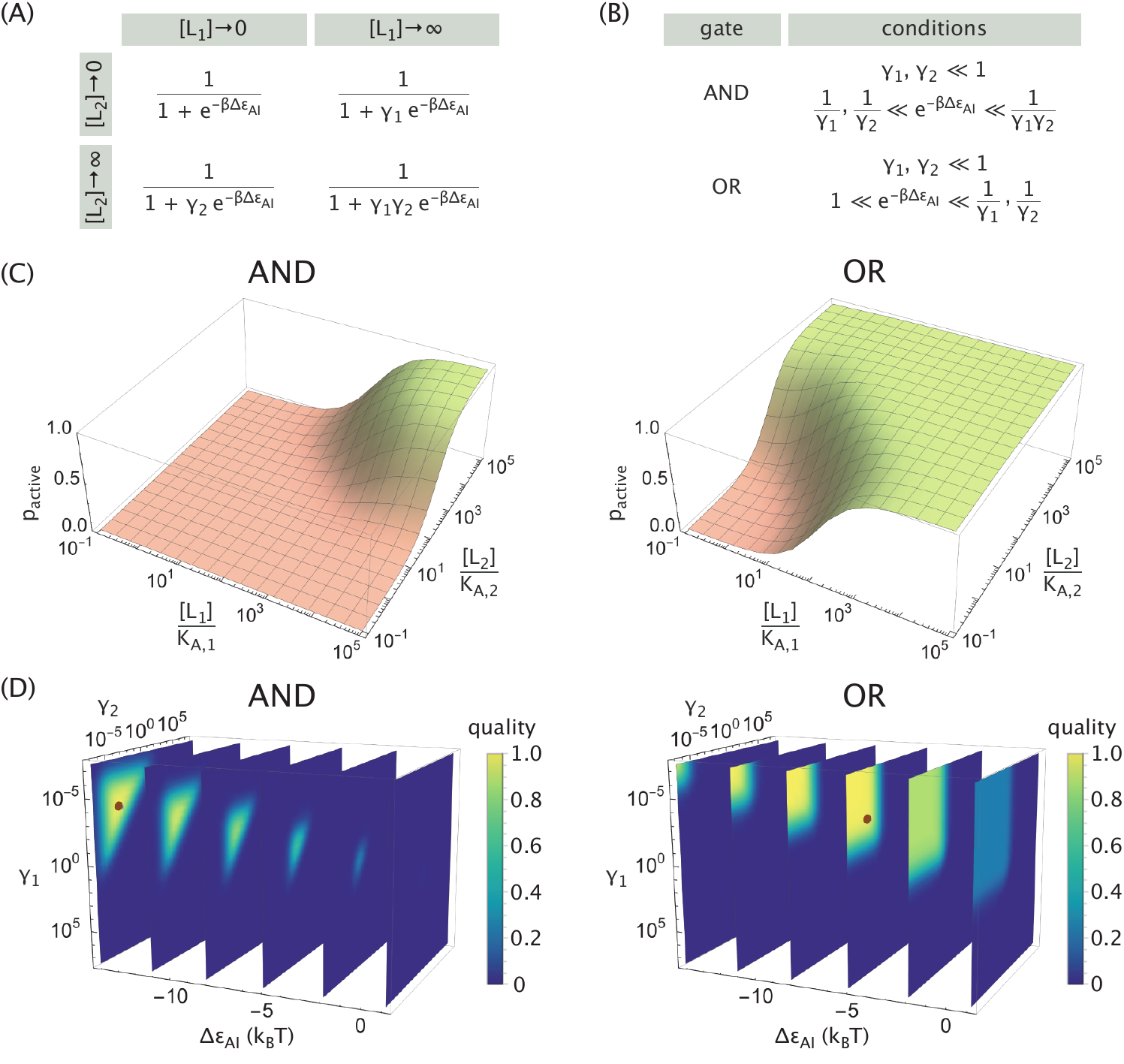}}
	\caption{\textbf{Logic gate realization of an allosteric protein with two
		ligands.} (A) Probability that the protein is active
	($\text{p}_{\text{active}}$) in different limits (rows and columns of the
	matrix) of ligand concentrations, where $\gamma_i =
	\frac{\text{K}_{\text{A},\text{i}}}{\text{K}_{\text{I},\text{i}}}$. (B)
	Conditions on the parameters that lead to an AND or OR response. (C)
	Realizations of the AND and OR logic gates. Parameters used were
	$\text{K}_{\text{A},1} = \text{K}_{\text{A},2} = 2.5 \times 10^{-8}\,\text{M}$,
	$\text{K}_{\text{I},1} = \text{K}_{\text{I},2} = 1.5 \times 10^{-4}\,\text{M}$,
	and $\Delta \text{\textepsilon}_{\text{AI}} = -14.2\,\text{k}_{\text{B}}
	\text{T}$ for the AND gate or $\Delta \text{\textepsilon}_{\text{AI}} =
	-5.0\,\text{k}_{\text{B}} \text{T}$ for the OR gate. (D) Quality of AND (eq~\ref{eq:AND_quality}) and OR (eq~\ref{eq:OR_quality})
	gates across parameter space. The brown dots indicate the high quality gates in
	Panel C.} \label{fig:MWC_simple}
\end{figure}

To determine whether this allosteric protein can serve as a molecular logic gate,
we first evaluate the probability that it is active when each ligand
is either absent ($[\text{L}_\text{i}] \rightarrow 0$) or at a saturating
concentration ($[\text{L}_\text{i}] \rightarrow \infty$).
Figure~\ref{fig:MWC_simple}A evaluates these limits for eq~\ref{eq:mwc}, where
we have introduced the parameters $\gamma_1 =
\frac{\text{K}_{\text{A},\text{1}}}{\text{K}_{\text{I},\text{1}}}$ and $\gamma_2
= \frac{\text{K}_{\text{A},\text{2}}}{\text{K}_{\text{I},\text{2}}}$ to simplify
the results.

The probabilities in Figure~\ref{fig:MWC_simple}A can be compared to the
target functions in Figure~\ref{fig:logic_gates} to determine the conditions on
each parameter that would be required to form a given logic gate. For example,
the AND, OR, and XOR gates require that in the absence of either ligand
($[\text{L}_1] = [\text{L}_2] = 0$), there should be as little activity as
possible, thereby requiring that the active state has a higher (more unfavored)
free energy than the inactive state ($\text{e}^{-\text{\textbeta} \Delta
	\text{\textepsilon}_{\text{AI}}} \gg 1$). We note that in the context of
transcriptional regulation, this limit of activity in the absence of ligands is
called the leakiness, \cite{Razo-Mejia2018} and it is one of the distinguishing features of the MWC
model in comparison with other allosteric models such as the Koshland-Némethy-Filmer (KNF)
model that exhibits no leakiness.

For the AND and OR gates, the condition that $\text{p}_{\text{active}} \approx
1$ when both ligands are saturating ($[\text{L}_1], [\text{L}_2] \rightarrow
\infty$) requires that $\gamma_1 \gamma_2 \text{e}^{-\text{\textbeta} \Delta
	\text{\textepsilon}_{\text{AI}}} \ll 1$. The two limits where one ligand is
absent while the other ligand is saturating lead to the conditions shown in
Figure~\ref{fig:MWC_simple}B for the AND and OR gates, with representative
response profiles shown in Figure~\ref{fig:MWC_simple}C using parameter values
from the single-ligand allosteric nicotinic acetylcholine receptor
\cite{AuerbachJPhysiol2012}. We relegate the derivations to
Appendix~\ref{Appendix_condition_derivations}, where we also demonstrate that
the XOR gate cannot be realized with the form of $\text{p}_{\text{active}}$ in
eq~\ref{eq:mwc} unless explicit cooperativity is added to the MWC model. In
addition, we show that the NAND, NOR, and XNOR gates can be formed if and only
if their complementary AND, OR, and XOR gates can be formed, respectively, by
replacing $\Delta\text{\textepsilon}_{\text{AI}} \to
-\Delta\text{\textepsilon}_{\text{AI}}$ and $\gamma_{\text{i}} \to
\frac{1}{\gamma_{\text{i}}}$. Finally, Figure~\ref{fig:MWC_simple}C
demonstrates that the same dissociation constants $\text{K}_{\text{A},\text{i}}$
and $\text{K}_{\text{I},\text{i}}$ can give rise to either AND or OR behavior by
modulating $\Delta\text{\textepsilon}_{\text{AI}}$, with the transition between
these two logic gates occurring at $\text{e}^{-\text{\textbeta} \Delta
	\text{\textepsilon}_{\text{AI}}} \approx \frac{1}{\gamma_1} \approx
\frac{1}{\gamma_2}$ (this corresponds to $\Delta\text{\textepsilon}_{\text{AI}}
\approx -9\,\text{k}_\text{B} \text{T}$ for the values of $\text{K}_{\text{A},\text{i}}$ 
and $\text{K}_{\text{I},\text{i}} $ in Figure~\ref{fig:MWC_simple}).

To explore the gating behavior changes across parameter space, we define a
quality metric for how closely $\text{p}_{\text{active}}$ matches its target
value at different concentration limits for a given idealized logic gate,
\begin{align} \label{eq:quality}
\text{Q}(\gamma_1, \gamma_2, \Delta \text{\textepsilon}_{\text{AI}}) &= \prod_{\lambda_1 \, = \, 0, \, \infty} \, \, \prod_{\lambda_2 \, = \, 0, \, \infty } (1-\left|\text{p}_{\lambda_1,\lambda_2}^\text{ideal} - \text{p}_{\lambda_1,\lambda_2} \right|),
\end{align}
where $\text{p}_{\lambda_1,\lambda_2} = \text{p}_{\text{active}}\left([\text{L}_1]\to \lambda_1,[\text{L}_2]\to\lambda_2\right)$. A value of 1 (high quality gate) implies a perfect match between the target
function and the behavior of the allosteric molecule while a value near 0 (low quality
gate) suggests that the response behavior deviates from the target function in
at least one limit.

From eq~\ref{eq:quality}, the quality for the AND gate becomes
\begin{equation} \label{eq:AND_quality}
\text{Q}_{\text{AND}} = (1-\text{p}_{0,0})(1-\text{p}_{\infty,0})(1-\text{p}_{0,\infty}) \text{p}_{\infty,\infty},
\end{equation}
while for the OR gate it takes on the form
\begin{equation} \label{eq:OR_quality}
\text{Q}_{\text{OR}} = (1-\text{p}_{0,0}) \, \text{p}_{\infty,0} \, \text{p}_{0,\infty} \, \text{p}_{\infty,\infty}.
\end{equation}
Figure~\ref{fig:MWC_simple}D shows the regions in parameter space where the
protein exhibits these gating behaviors (the high quality gates from
Figure~\ref{fig:MWC_simple}C are denoted by brown dots). More specifically,
for a fixed $\Delta \text{\textepsilon}_{\text{AI}}$, the AND behavior is
achieved in a finite triangular region in the $\gamma_1$-$\gamma_2$ plane which
grows larger as $\Delta \text{\textepsilon}_{\text{AI}}$ decreases. The OR gate,
on the other hand, is achieved in an infinite region defined by $\gamma_1, \gamma_2
\lesssim \text{e}^{\text{\textbeta} \Delta \text{\textepsilon}_{\text{AI}}}$.
In either case, a high quality gate can be obtained only when 
the base activity is very low 
($\Delta \text{\textepsilon}_{\text{AI}} \lesssim 0$)
and when both ligands are strong activators ($\gamma_1, \gamma_2 \ll 1$),
in agreement with the derived conditions (Figure~\ref{fig:MWC_simple}B).
Lastly, we note that the quality
metrics for AND/OR and their complementary NAND/NOR gates obey a simple
relation, namely, $\text{Q}_{\text{AND/OR}} \left( \gamma_1, \gamma_2, \Delta
\text{\textepsilon}_{\text{AI}} \right) = \text{Q}_{\text{NAND/NOR}} \left(
\frac{1}{\gamma_1}, \frac{1}{\gamma_2}, -\Delta \text{\textepsilon}_{\text{AI}}
\right)$, which follows from the functional form of eq~\ref{eq:quality} and the
symmetry between the two gates (see
Appendix~\ref{Appendix_condition_derivations}).

\subsection*{General Two-Ligand MWC Response}

\begin{figure}[!ht]
	\centerline{\includegraphics{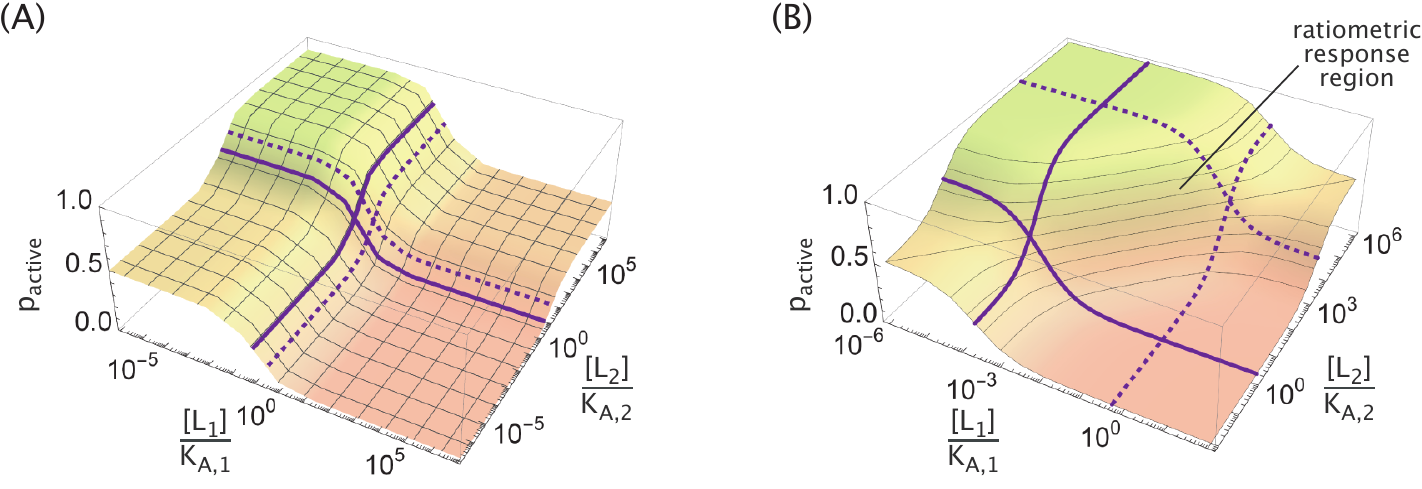}}
	\caption{\textbf{General MWC response with two ligands.} (A) Three of the four
	limits of ligand concentrations ($[\text{L}_1], [\text{L}_2] \to 0$ or
	$\infty$) can be fixed by the parameters $\Delta
	\text{\textepsilon}_{\text{AI}}$, $\gamma_\text{1}$, and $\gamma_\text{2}$.
	Additionally, the midpoint of the $[\text{L}_\text{i}]$ response when
	$[\text{L}_\text{j}] \to 0$ (solid purple curve) or $[\text{L}_\text{j}] \to
	\infty$ (dashed purple curve) can be adjusted. (B) Within the region determined
	by the four midpoints, the MWC response becomes ratiometric
	\cite{Antebi2017} where the concentration ratio of the two ligands
	determines the activity of the molecule.
	This is illustrated by the diagonal contour lines of constant $\text{p}_{\text{active}}$
	in the ratiometric response region.}
	\label{fig:general_two_ligand_response}
\end{figure}

We next relax the constraint that $\text{p}_{\text{active}}$ must either
approach 0 or 1 in the limits of no ligand or saturating ligand and consider the
general behavior that can be achieved by an MWC molecule in the four limits
shown in Figure~\ref{fig:MWC_simple}A. Manipulating the three parameters
($\gamma_1$, $\gamma_2$ and $\Delta \text{\textepsilon}_{\text{AI}}$) enables us
to fix three of the four limits of $\text{p}_{\text{active}}$, and these three
choices determine the remaining limit. For example, the parameters in
Figure~\ref{fig:general_two_ligand_response}A were chosen so that
$\text{p}_{0,0} = 0.5$ ($\Delta \text{\textepsilon}_{\text{AI}} = 0$),
$\text{p}_{0,\infty} \approx 0.9$ ($\gamma_2 = 0.1$), and $\text{p}_{\infty,0} \approx 0.05$
($\gamma_1 = 20$), which fixed $\text{p}_{\infty,\infty} \approx 0.3$ for the final
limit.

In addition to the limits of $\text{p}_{\text{active}}$, the locations of the
transitions between these limits can be controlled by changing
$\text{K}_{\text{A},\text{i}}$ and $\text{K}_{\text{I},\text{i}}$ while
keeping $\gamma_{\text{i}} =
\frac{\text{K}_{\text{A},\text{i}}}{\text{K}_{\text{I},\text{i}}}$ constant. In
Appendix~\ref{Appendix_general_response} we generalize previous results for the
transition of a single-ligand MWC receptor \cite{MarzenJMolBio2013} to the
present case of two ligands. Interestingly, we find that the midpoint
$[\text{L}_{1}^*]_{[\text{L}_{2}] \to 0}$ of the response
in the absence of $[\text{L}_{2}]$
(solid curve in Figure~\ref{fig:general_two_ligand_response}A) is
different from the midpoint $[\text{L}_{1}^*]_{[\text{L}_{2}] \to \infty}$ of the response at
saturating $[\text{L}_{2}]$ (dashed curve in
Figure~\ref{fig:general_two_ligand_response}A), 
with analogous statements holding for the second ligand. More precisely, the two transition points occur at
\begin{align}
	[\text{L}_{\text{i}}^*]_{{[\text{L}_{\text{j}}] \to 0}} &= \text{K}_{\text{A},\text{i}} \frac{1+\text{e}^{-\text{\textbeta} \Delta
	\text{\textepsilon}_{\text{AI}}}}{1+\gamma_\text{i} \, \text{e}^{-\text{\textbeta} \Delta
	\text{\textepsilon}_{\text{AI}}}}, \label{eq:transition_concentration_1_absent} \\
	[\text{L}_{\text{i}}^*]_{{[\text{L}_{\text{j}}] \to \infty}} &= \text{K}_{\text{A},\text{i}} \frac{1 + \gamma_\text{j} \, \text{e}^{-\text{\textbeta} \Delta
	\text{\textepsilon}_{\text{AI}}}}{1+\gamma_1 \gamma_2 \, \text{e}^{-\text{\textbeta} \Delta
	\text{\textepsilon}_{\text{AI}}}}. \label{eq:transition_concentration_1_saturating}
\end{align}
Notably, the ratio
\begin{align}\label{eq:transition_ratio}
\frac{[\text{L}_{\text{i}}^*]_{{[\text{L}_{\text{j}}] \to \infty}}}{[\text{L}_{\text{i}}^*]_{{[\text{L}_{\text{j}}] \to 0}}} = \frac{(1 + \gamma_1 \, \text{e}^{-\text{\textbeta} \Delta
	\text{\textepsilon}_{\text{AI}}})(1 + \gamma_2 \,\text{e}^{-\text{\textbeta} \Delta
	\text{\textepsilon}_{\text{AI}}})}{(1+\text{e}^{-\text{\textbeta} \Delta
	\text{\textepsilon}_{\text{AI}}})(1+\gamma_1 \gamma_2 \,\text{e}^{-\text{\textbeta} \Delta
	\text{\textepsilon}_{\text{AI}}})}
\end{align}
is invariant to ligand swapping ($\text{i}\leftrightarrow\text{j}$); hence, the transition zones, 
defined as the concentration intervals between solid and dotted curves, have identical sizes for the 
two ligands, as can be seen in Figure~\ref{fig:general_two_ligand_response}.

The MWC response has its steepest slope when the ligand concentration is within
the range set by $[\text{L}_{\text{i}}^*]_{{[\text{L}_{\text{j}}] \to 0}}$ and
$[\text{L}_{\text{i}}^*]_{{[\text{L}_{\text{j}}] \to \infty}}$, 
and interesting response behaviors can arise when both
ligand concentrations fall into this regime. 
For example, Antebi \textit{et
	al.}~recently showed that the BMP pathway exhibits ratiometric response where 
pathway activity depends monotonically on the ratio of the ligand concentrations.\cite{Antebi2017} Similar response functions 
have also been observed in the GAL pathway in yeast, 
where gene induction is sensitive to the ratio of galactose and glucose.\cite{EscalanteChong2015}
Such behavior can be 
achieved within the highly sensitive region of the MWC
model using one repressor ligand ($\text{L}_1$) and one
activator ligand ($\text{L}_2$), as shown in Figure~\ref{fig:general_two_ligand_response}B.
Parameters
chosen for demonstration are 
$\Delta \text{\textepsilon}_{\text{AI}} = 0$, $\text{K}_{\text{A},\text{1}} =
\text{K}_{\text{A},\text{2}}$ and 
$\frac{\text{K}_{\text{I},\text{1}}}{\text{K}_{\text{A},\text{1}}} =
\frac{\text{K}_{\text{A},\text{2}}}{\text{K}_{\text{I},\text{2}}} =10^{-4}$.
In this regime, the 
probability of the protein being active gets reduced to
\begin{equation}
\label{eq:ratiometric}
\text{p}_{\text{active}}\left( [\text{L}_1], [\text{L}_2] \right) \approx \frac{\frac{[\text{L}_2]}{\text{K}_{\text{A},2}}}{\frac{[\text{L}_2]}{\text{K}_{\text{A},2}} + \frac{[\text{L}_1]}{\text{K}_{\text{I},1}}},
\end{equation}
which clearly depends monotonically on the $[\text{L}_2]/[\text{L}_1]$ ratio
(see Appendix~\ref{Appendix_general_response} for details).
We note that the region over which the ratiometric behavior is observed 
can be made arbitrarily large by decreasing the ratios
$\frac{\text{K}_{\text{I},\text{1}}}{\text{K}_{\text{A},\text{1}}}$ and
$\frac{\text{K}_{\text{A},\text{2}}}{\text{K}_{\text{I},\text{2}}}$.

\subsection*{Modulation by Multiple Ligands}

A much richer repertoire of signaling responses is available to an MWC protein
if we go beyond two ligand inputs with a single binding site for each, 
as exhibited by phosphofructokinase, for example. 
Though earlier we mentioned phosphofructokinase in the context of two of its input ligands,
in fact, this enzyme has even more inputs than that and thus provides a rich example of multi-ligand
combinatorial control. \cite{Blangy1968}
To start exploring the diversity of these responses, we generalize eq~\ref{eq:mwc} to
consider cases with $\text{N}$ input ligands, where the $\text{i}^\text{th}$
ligand has $\text{n}_\text{i}$ binding sites, concentration
$[\text{L}_\text{i}]$, and dissociation constants $\text{K}_{\text{A},\text{i}}$
and $\text{K}_{\text{I},\text{i}}$ with the molecule's active and inactive
states, respectively. In general, it is impractical to write the states and
weights as we have done in Figure~\ref{fig:states_weights}, since the total
number of possible states, given by $2^{1 + \sum_{\text{i}=1}^{\text{N}}
	\text{n}_\text{i}}$, grows exponentially with the number of binding sites.
However, by analogy with the earlier simple case,  the general formula for the
probability that the protein is active can be written as
\begin{equation}
\label{eq:mwc_generalize}
\text{p}_{\text{active}} \left( [\text{L}_1], [\text{L}_2], ..., [\text{L}_\text{N}] \right) = \frac{ \prod_{\text{i}=1}^{\text{N}}\left( 1 + \frac{[\text{L}_\text{i}]}{\text{K}_{\text{A},\text{i}}} \right)^{\text{n}_\text{i}}}{\prod_{\text{i}=1}^{\text{N}}\left( 1 + \frac{[\text{L}_\text{i}]}{\text{K}_{\text{A},\text{i}}} \right)^{\text{n}_\text{i}}+ \text{e}^{-\text{\textbeta} \Delta \text{\textepsilon}_{\text{AI}}} \prod_{\text{i}=1}^{\text{N}}\left( 1 + \frac{[\text{L}_\text{i}]}{\text{K}_{\text{I},\text{i}}} \right)^{\text{n}_\text{i}}}.
\end{equation}

We first consider an MWC molecule with $\text{N}=2$ input ligands as in the
previous section but with $\text{n}_\text{i}$ ligand binding sites for ligand
$\text{i}$. As derived in Appendix~\ref{Appendix_increased_binding_sites}, the
criteria for the AND and OR gates are identical to those for a protein with
$\text{n}_\text{i}=1$ binding site per ligand, except that we make the
$\gamma_\text{i} \to \gamma_\text{i}^{\text{n}_\text{i}}$ substitution in the
conditions shown in Figure~\ref{fig:MWC_simple}B. The protein thus exhibits OR
behavior if $\text{e}^{-\text{\textbeta} \Delta \text{\textepsilon}_{\text{AI}}}
\ll \min \left( \frac{1}{\gamma_1^{\text{n}_1}}, \frac{1}{\gamma_2^{\text{n}_2}}
\right) $ or AND behavior if $\text{e}^{-\text{\textbeta} \Delta
	\text{\textepsilon}_{\text{AI}}} \gg \max \left(
\frac{1}{\gamma_1^{\text{n}_1}}, \frac{1}{\gamma_2^{\text{n}_2}} \right)$.

\begin{figure}[!ht]
	\centerline{\includegraphics{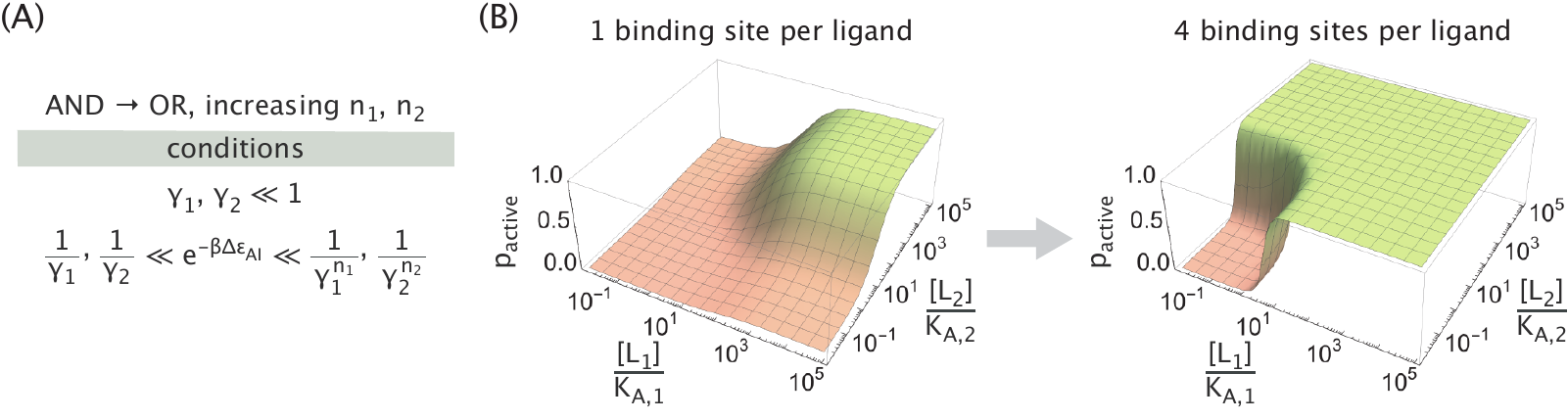}}
	\caption{\textbf{Increased number of binding sites can switch the logic of an MWC protein
			from AND into OR.}
		(A) Parameter conditions required for $\text{AND}\to\text{OR}$ switching
		upon an increase in the number of binding sites.
		(B) Representative activity plots showing the $\text{AND}\to\text{OR}$ switching.
		Parameters used were
		$\text{K}_{\text{A},\text{i}} = 2.5 \times 10^{-8}\, \text{M}$,
		$\text{K}_{\text{I},\text{i}} = 2.5 \times 10^{-6}\, \text{M}$ and $\Delta
		\text{\textepsilon}_{\text{AI}} = -7\,\text{k}_{\text{B}} \text{T}$.}
	\label{fig:mwc_generalizations}
\end{figure}

Over evolutionary time or through synthetic approaches, 
the number of binding sites displayed by a single molecule 
can be tuned, enabling such
systems to test a variety of responses with a limited repertoire of regulatory
molecules. Since $\gamma_1,\gamma_2 \ll 1$, increasing the number of binding sites while
keeping all other parameters the same can shift a response from AND$\to$OR as
shown in Figure~\ref{fig:mwc_generalizations}. The opposite logic switching
(OR$\rightarrow$AND) is similarly possible by decreasing the number of binding
sites, and analogous results can be derived for the complementary NAND and NOR
gates (see Appendix~\ref{Appendix_increased_binding_sites}). In the limit where
the number of binding sites becomes large ($\text{n}_1, \text{n}_2 \gg 1$), an
allosteric molecule's behavior will necessarily collapse into OR logic provided
$\gamma_1,\gamma_2 < 1$, since the presence of either ligand occupying the
numerous binding sites has sufficient free energy to overcome the
active-inactive free energy difference $\Delta \text{\textepsilon}_{\text{AI}}$.
In addition, having a large number of binding sites makes the
$\text{p}_{\text{active}}$ response sharper
(Figure~\ref{fig:mwc_generalizations}B), as has been seen in the context of
chromatin remodeling where $\sim$150 bp of DNA ``buried'' within a nucleosome
can be made available for transcription by the binding of multiple transcription
factors. \cite{MirnyPNAS2010}

\begin{figure}[!ht]
	\centerline{\includegraphics{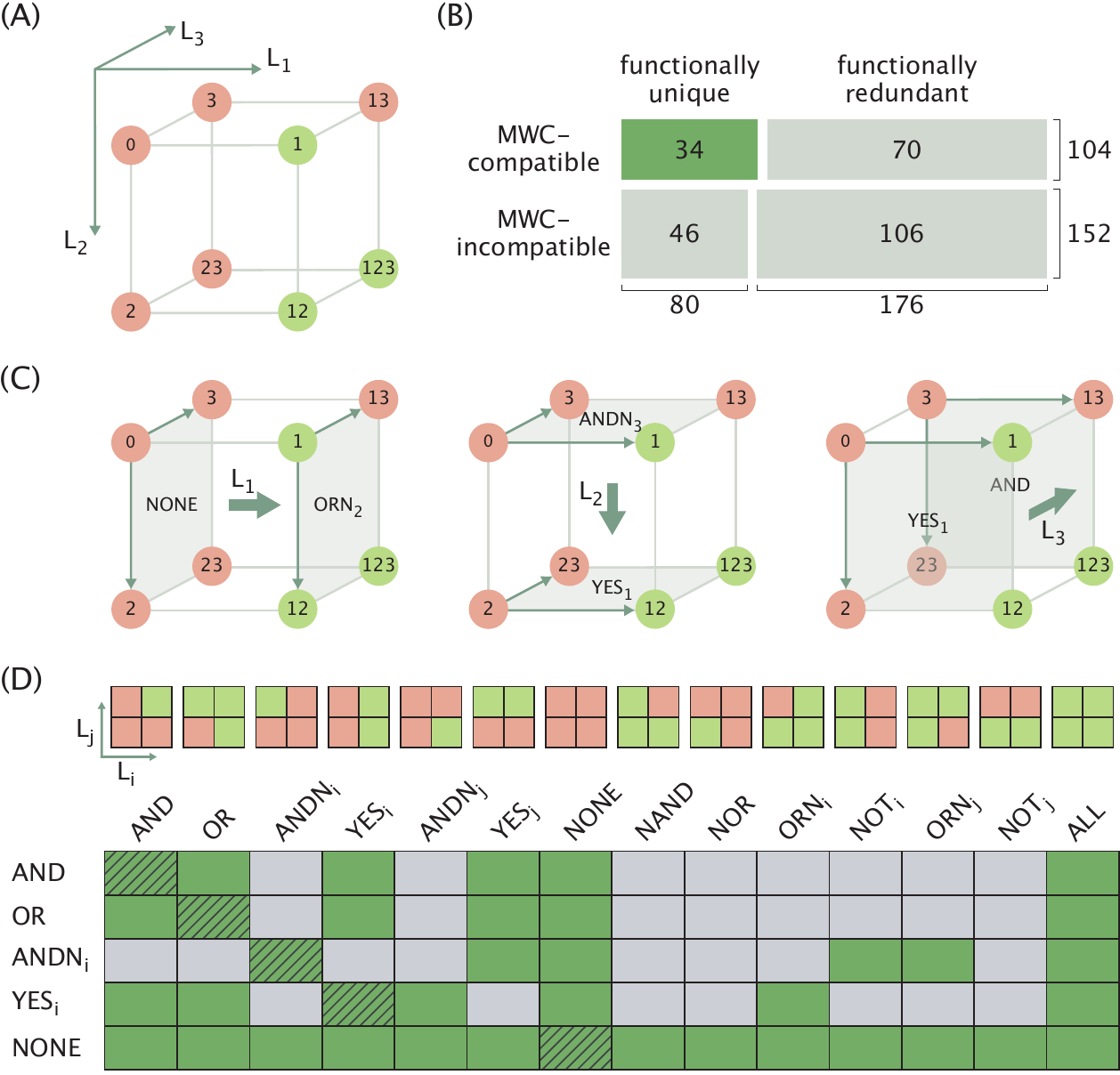}}
	\caption{\textbf{Third ligand expands the combinatorial diversity of logic
			responses and enables logic switching.} 
		(A) Cubic diagram of a representative molecular logic response. The label ``0''
		stands for the limit when all ligands are at low concentrations. Each digit in
		the labels of other limits indicates the high concentration of the
		corresponding ligand (for example, in the ``12'' limit the ligands 1 and 2 are
		at high concentrations). Red and green colors indicate the OFF and ON states of
		the molecule, respectively.
		(B) Diagram representing the numbers of 3-ligand logic gates categorized
		by their MWC compatibility and functional uniqueness.
		The area of each cell is proportional to the number of gates in the corresponding category.
		(C) Demonstration of different logic transitions induced by a third ligand (thick arrows)
		on the example of the 3-input gate in Panel A.
		(D) Table of all possible logic transitions (row
		$\to$ column, green cells) inducible by a third ligand in the MWC
		framework. Schematics of the 14 MWC-compatible 2-ligand gates 
		corresponding to each column entry are displayed 
		on top ($\text{i}$ and $\text{j}$ represent different ligands). 
		Results for the transitions between logical complements (NOT row 
		$\to$ NOT column) are identical to the results for row  $\to$
		column transitions and are not shown.
		Trivial transitions between identical gates where the third ligand
		has no effect are marked with hatching lines.}
	\label{fig:ThreeLigandPlusTransitions}
\end{figure}

Next, we examine an alternative possibility of generalizing the MWC response,
namely, considering a molecule with $\text{N}=3$ distinct ligands, each having a
single binding site ($\text{n}_\text{i}=1$).
The logic response is now described
by a $2\times2\times2$ cube corresponding to the activity at low and
saturating concentrations of each of the three ligands (an example realization
is shown in Figure \ref{fig:ThreeLigandPlusTransitions}A). Since each of the 8 cube
elements can be either OFF or ON (red and green circles, respectively), the
total number of possible responses becomes $2^8 = 256$. This number, however,
includes functionally redundant responses, as well as ones that are not
admissible in the MWC framework. We therefore eliminate these cases in order to 
accurately quantify the functional diversity of 3-input MWC proteins.

We consider two responses to be functionally identical if one can be obtained
from another by relabeling the ligands, e.g. $(1,2,3) \to (3,1,2)$. Eliminating
all redundant responses leaves 80 unique cases out of the 256 possibilities (see
Appendix \ref{Appendx_three_ligands}). In addition, since the molecule's activity
in the eight ligand concentration limits is determined by only four MWC
parameters, namely, $\{ \Delta \text{\textepsilon}_{\text{AI}}, \gamma_1,
\gamma_2, \gamma_3 \}$, we expect the space of possible 3-input gates to be
constrained (analogous to XOR/XNOR gates being inaccessible to 2-input MWC
proteins). Imposing the constraints leaves 34 functionally unique logic
responses that are compatible with the MWC framework (see Figure
\ref{fig:ThreeLigandPlusTransitions}B for the summary statistics and Appendix
\ref{Appendx_three_ligands} for the detailed discussion of how the constraints
were imposed).

In addition to expanding the scope of combinatorial control relative to the
two-input case, we can think of the role of the third ligand as a regulator
whose presence switches the logic performed by the other two ligands. We
illustrate this role in Figure~\ref{fig:ThreeLigandPlusTransitions}C by first
focusing on the leftmost cubic diagram. The gating behavior on the left face of the cube (in
the absence of $\text{L}_1$) exhibits NONE logic while the behavior on the right
face of the cube (in the presence of saturating $\text{L}_1$) is the
$\text{ORN}_2$ logic (see the schematics at the top of
Figure~\ref{fig:ThreeLigandPlusTransitions}D for the definition of all possible gates). In this
way, adding $\text{L}_1$ switches the logic of the remaining two ligands from
$\text{NONE} \to \text{ORN}_2$. In a similar vein, adding $\text{L}_2$ changes
the logic from $\text{ANDN}_3 \to \text{YES}_1$, while adding $\text{L}_3$
causes a $\text{YES}_1 \to \text{AND}$ switch.

We repeat the same procedure for all functionally unique 3-ligand MWC gates (see
Appendix \ref{Appendx_three_ligands}) and obtain a table of all possible logic
switches that can be induced by a third ligand (green cells in Figure
\ref{fig:ThreeLigandPlusTransitions}D that indicate $\text{row}\to\text{column}$ logic switches).
As we can see, a large set of logic
switches are feasible, the majority of which (the left half of the table) do not
involve a change in the base activity (i.e., activity in the absence of
the two ligands). Comparatively fewer transitions that involve flipping of the
base activity from OFF to ON are possible (the right half of the
table).

\begin{figure}[!ht]
	\centerline{\includegraphics{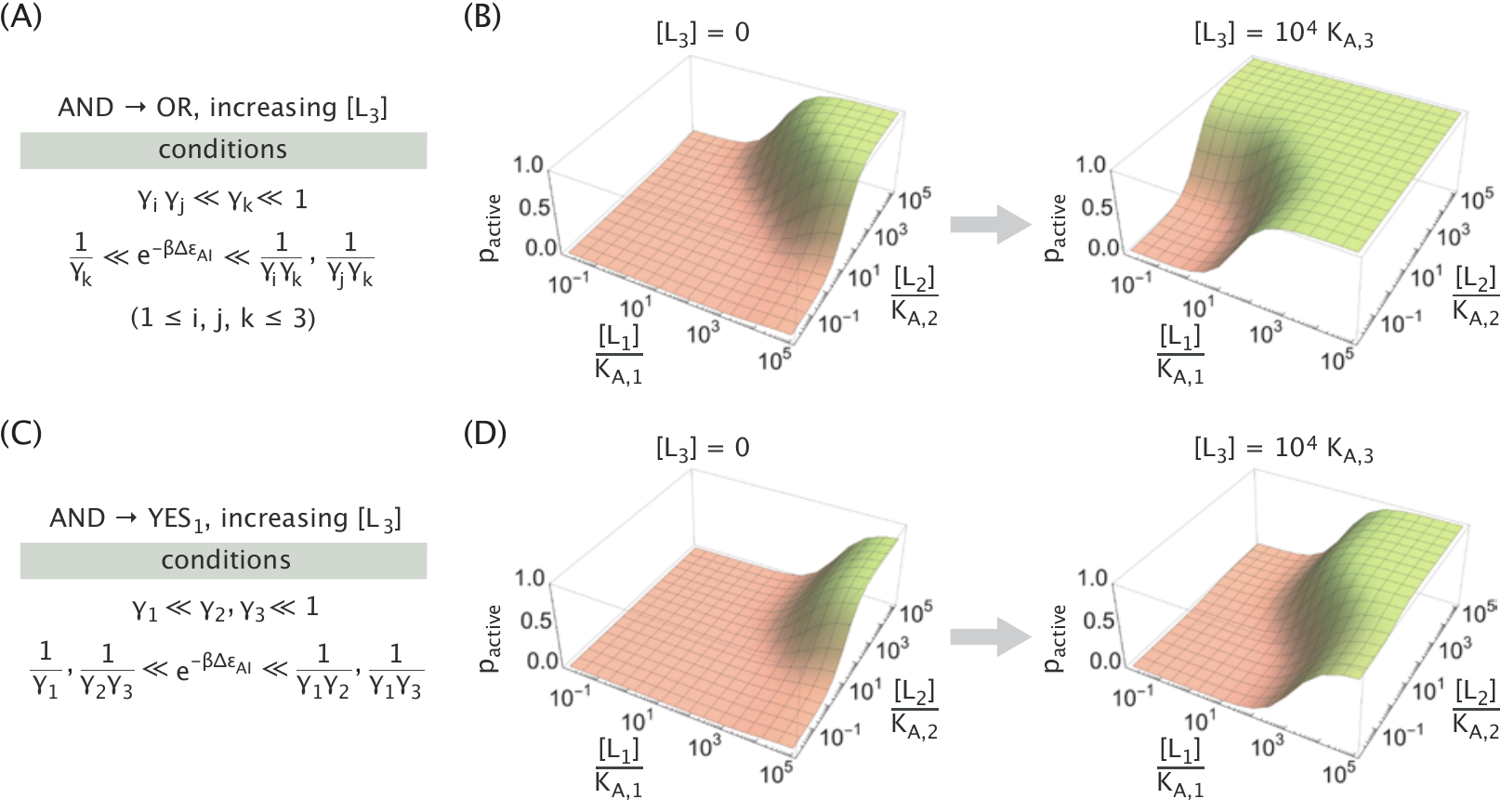}}
	\caption{\textbf{Example logic switches induced by the third ligand.} Parameter
		conditions and representative activity plots of an allosteric molecule exhibiting AND logic
		in the absence of the third ligand, while exhibiting OR logic (A,B) or
		$\text{YES}_1$ logic (C,D) when $\text{L}_3$ is present at a saturating
		concentration. Parameters used were $\text{K}_{\text{A},\text{i}} = 2.5 \times
		10^{-8}\, \text{M}$ and $\text{K}_{\text{I},\text{i}} = 2.5 \times 10^{-4}\,
		\text{M}$ in Panel B,  $\text{K}_{\text{A},\text{i}} = 2.5 \times 10^{-8}\,
		\text{M}$, $\text{K}_{\text{I},\text{1}} = 2.5 \times 10^{-4}\, \text{M}$ and
		$\text{K}_{\text{I},\text{2/3}} = 2.5 \times 10^{-6}\, \text{M}$ in panel D,
		along with $\Delta \text{\textepsilon}_{\text{AI}} = -12\,\text{k}_{\text{B}}
		\text{T}$ in both panels.} \label{fig:exampleSwitches}
\end{figure}

As a demonstration of the regulatory function of the third ligand, we show two
examples of logic switching induced by increasing [$\text{L}_3$], namely,
AND$\rightarrow$OR (Figure \ref{fig:exampleSwitches}A,B) and
AND$\rightarrow$YES$_{\text{1}}$ (Figure \ref{fig:exampleSwitches}C,D), along
with the parameter conditions that need to be satisfied to enable such
transitions (see Appendix \ref{Appendx_three_ligands} for derivations). An
interesting perspective is to view the $\text{L}_3$ ligand as a modulator of the
free energy difference $\Delta \text{\textepsilon}_{\text{AI}}$. For example,
when $[\text{L}_3] = 0$, the protein behaves identically to the $\text{N}=2$
case given by eq~\ref{eq:mwc}; at a saturating concentration of $\text{L}_3$,
however, the protein behaves as if it had $\text{N}=2$ ligands with a modified
free energy difference $\Delta\text{\textepsilon}^\prime_{\text{AI}}$ given by
\begin{align}
\Delta\text{\textepsilon}^\prime_{\text{AI}} =
\Delta\text{\textepsilon}_{\text{AI}} -  \text{k}_\text{B} \text{T} \log \gamma_3.
\end{align}
From this perspective, the third ligand increases the effective free energy
difference in the examples shown in Figure \ref{fig:exampleSwitches}, since in
both cases the $\gamma_3\ll1$ condition is satisfied. For the AND$\rightarrow$OR
transition, the increase in $\Delta\text{\textepsilon}_{\text{AI}}$ is
sufficient to let either of the two ligands activate the molecule (hence, the OR
gate). In the {AND$\rightarrow$YES$_1$} transition, the change in
$\Delta\text{\textepsilon}_{\text{AI}}$ utilizes the asymmetry between the
binding strengths of the two ligands ($\gamma_1 \ll \gamma_2$) to effectively
``silence'' the activity of the ligand $\text{L}_2$. We note in passing that
such behavior for the $\text{N}=3$ allosteric molecule is reminiscent of a transistor which
can switch an input signal in electronics.

\section*{Discussion and Conclusions}

Combinatorial control is a ubiquitous strategy employed by cells. Networks of
cellular systems of different kinds, such as transcriptional, \cite{Scholes2017,
	KinkhabwalaGuetPloSOne2008} signaling, \cite{DueberLimCurrOpStructBiol2004} or
metabolic, \cite{Blangy1968} integrate information from multiple inputs in order
to produce a single output. The statistical mechanical MWC model we employ
allows us to systematically explore the combinatorial diversity of output
responses available to such networks and determine the conditions that the MWC
parameters need to satisfy to realize a particular response.

In this paper, we built on earlier work to show that the response of an allosteric MWC molecule 
can mimic Boolean logic.
Specifically, we
demonstrated that a protein that binds to two ligands can
exhibit an AND, OR, NAND, or NOR response (also shown by 
others \cite{Graham2005,deRonde2012,Agliari2015}), where the former two cases require the
protein to be inherently inactive and that both ligands preferentially bind to
the active conformation, whereas the latter two cases require the converse
conditions.
We derived the MWC parameter ranges within which an allosteric protein
would exhibit an AND or OR response (Figure \ref{fig:MWC_simple}B), and showed 
that the corresponding 
parameter ranges for NAND or NOR responses could be achieved by
simply substituting $\gamma_\text{i} \to \frac{1}{\gamma_\text{i}}$ 
and $\Delta \text{\textepsilon}_{\text{AI}} \to - \Delta \text{\textepsilon}_{\text{AI}}$
in the parameter condition equations (Appendix \ref{appendix:NANDandNOR}).
Since the NAND and NOR gates are known in digital electronics as
universal logic gates, all other logic functions can be reproduced by
hierarchically layering these gates. In the context of this work, such layering
could be implemented if the MWC protein is an enzyme that only catalyzes in the
active state so that its output (the amount of product) could serve as an input
for the next enzyme, thereby producing more complex logic functions via
allostery, though at the cost of noise amplification and response delays.

As in earlier work,\cite{Graham2005,deRonde2012} we showed that 
the XOR and XNOR responses cannot be achieved
within the original MWC framework (eq~\ref{eq:mwc}) but are possible when
cooperativity between the two ligands is introduced
(Appendix~\ref{Appendix_XOR}).
Biological XOR and XNOR behaviors are uncommon in 
non-transcriptional systems and have also 
been challenging for synthetic design and optimization.
\cite{PrivmanKatzJPhysChemB2010}.
One of the few examples of such systems is a synthetic metallochromic
chromophore whose transmittance output level is modulated
by $\text{Ca}^{2+}$ and $\text{H}^+$ ions in a XOR-like manner.
\cite{deSilvaMcClenaghanChemEuroJ2002, deSilvaUchiyamaNatNanotech2007}.

In addition to traditional Boolean logic, we recognized further manifestations of 
combinatorial control by two-ligand MWC proteins.
In particular, we showed that the protein activity
in three of the four ligand concentration limits can be set independently by tuning
the MWC parameters $\gamma_1$, $\gamma_2$, and $\Delta\text{\textepsilon}_{\text{AI}}$,
and that the ligand concentrations at which transitions between limit responses take place
can be separately controlled by proportionally changing 
$\text{K}_{\text{A},\text{i}}$ and $\text{K}_{\text{I},\text{i}}$, while
keeping $\gamma_{\text{i}} = \frac{\text{K}_{\text{A},\text{i}}}{\text{K}_{\text{I},\text{i}}}$ constant
(eqs~\ref{eq:transition_concentration_1_absent} and \ref{eq:transition_concentration_1_saturating}).
We also showed that when the ranges of ligand concentrations are close to
those transition values, then ratiometric sensing observed in the 
BMP \cite{Antebi2017} and  GAL pathways, 
\cite{EscalanteChong2015} can be recapitulated through
the MWC model (Figure \ref{fig:general_two_ligand_response}B), 
with larger regions of sensitivity achievable
by an appropriate tuning of the parameters.
We note that parameter ``tuning'' can be realized 
either through evolutionary processes over long time scales or 
synthetically, using mutagenesis or other approaches.
\cite{BloomArnoldCurrOpStructBiol2005}

Apart from altering the thermodynamic parameters such as the ligand
binding affinity or the free energy of active and inactive protein conformations,
the number of ligand binding sites of an allosteric molecule can also be changed.
This can occur evolutionarily 
through recombination events, synthetically by engineering combinations of protein domains,
\cite{GuntasOstermeierJMolBiol2004} or through binding of competitive effectors that
reduce the effective number of ligand binding sites.
We found that these alterations in the number of
ligand binding sites are capable of switching the logic behavior
between AND$\leftrightarrow$OR or NAND$\leftrightarrow$NOR gates
(Figure~\ref{fig:mwc_generalizations}B).
Since the MWC model has even been applied in unusual situations 
such as the packing of DNA into nucleosomes,\cite{MirnyPNAS2010, NarulaIgoshinIETSysBiol2010}
these results on combinatorial control can also be relevant for eukaryotic transcription.
The opening of the nucleosome is itself often subject to combinatorial control because there can be
multiple transcription factor binding sites within a given nucleosome, the number of which can also be tuned using
synthetic approaches. \cite{Lohr2009, Fakhouri2010,Chen2012, Crocker2017}

Lastly, we generalized the analysis of logic responses for a molecule whose
activity is modulated by three ligands, and identified 34 functionally unique
and MWC-compatible gates out of 256 total possibilities. 
We offered a perspective
on the function of any of the three ligands as a ``regulator'' that 
can cause a switch in the type of logic performed by the other two ligands
and derived the full list of such switches (Figure \ref{fig:ThreeLigandPlusTransitions}D).
Within the MWC model, the role of this regulatory ligand can be viewed as
effectively changing the free energy difference $\Delta\text{\textepsilon}_{\text{AI}}$
between the protein's active and inactive states (Appendix \ref{appendix:logic_switching}),
which, in turn, is 
akin to the role of methylation 
\cite{HansenWingreenPLoSCompBiol2008, LanTuNatPhys2012} or 
phosphorylation \cite{LanTuNatPhys2012} in adaptation, but without the
covalent linkage.
Our in-depth analysis of the logic repertoire available to 3-input MWC molecules
can serve as a theoretical framework for designing new allosteric proteins and also
for understanding the measured responses of existing systems.
Examples of such systems that both act as 3-input AND gates include
the GIRK channel, the 
state of which (open or closed) is regulated by the G protein 
$\text{G}_{\text{\textbeta} \gamma}$, the lipid $\text{PIP}_2$ and $\text{Na}^+$ ions
\cite{WangMacKinnonELife2016}, 
or the engineered N-WASP signaling protein which is activated by 
SH3, Cdc42 and PDZ ligands. \cite{Dueber2007}

The exquisite control that arises from the web of interactions underlying
biological systems is difficult to understand and replicate. A first step to
overcoming this hurdle is to carefully quantify the types of behaviors that can
arise from multi-component systems. As our ability to harness and potentially
design \textit{de novo} allosteric systems grows \cite{Raman2016, WangMacKinnonELife2016,
	HuangNature2016, GuntasOstermeierJMolBiol2004, Dueber2007, GuntasOstermeierPNAS2005}, 
	we can augment our current level of combinatorial control in
biological contexts, such as transcriptional regulation, \cite{Buchler2003,
	Scholes2017, WeiGuanSciRep2016, MaciaSole2012, KinkhabwalaGuetPloSOne2008} 
	to create even richer dynamics.


\section*{Acknowledgements}

It is a great pleasure to acknowledge the contributions of Bill Eaton to our 
understanding of allostery.
We thank Chandana Gopalakrishnappa and Parijat Sil for their input on this work, 
and Michael Elowitz for his insights and valuable feedback on the manuscript.
This research was supported by La Fondation Pierre-Gilles de Gennes, the Rosen
Center at Caltech, the Department of Defense through the National Defense Science 
\& Engineering Graduate Fellowship (NDSEG) Program (LF), and the National Institutes of 
Health DP1 OD000217 (Director’s Pioneer Award), R01 GM085286, and 1R35 GM118043-01 (MIRA). 
We are grateful to the Burroughs-Wellcome Fund for its support of the Physical Biology
of the Cell Course at the Marine Biological Laboratory, where part of this work was completed.

\bibliography{PaperLibrary}


\pagebreak
\appendix

\setcounter{equation}{0}
\setcounter{figure}{0}
\setcounter{page}{1}
\renewcommand{\thepage}{S\arabic{page}}
\renewcommand{\thefigure}{S\arabic{figure}}
\renewcommand{\thetable}{S\arabic{table}}
\renewcommand{\theequation}{S\arabic{equation}}

\newpage


\renewcommand\contentsname{Supporting Information Contents}
\tableofcontents
\addtocontents{toc}{\protect\setcounter{tocdepth}{2}}

\newpage
\section{Derivation of Conditions for Achieving Different Logic Responses} \label{Appendix_condition_derivations}
\begin{figure}[!ht]
	\centerline{\includegraphics{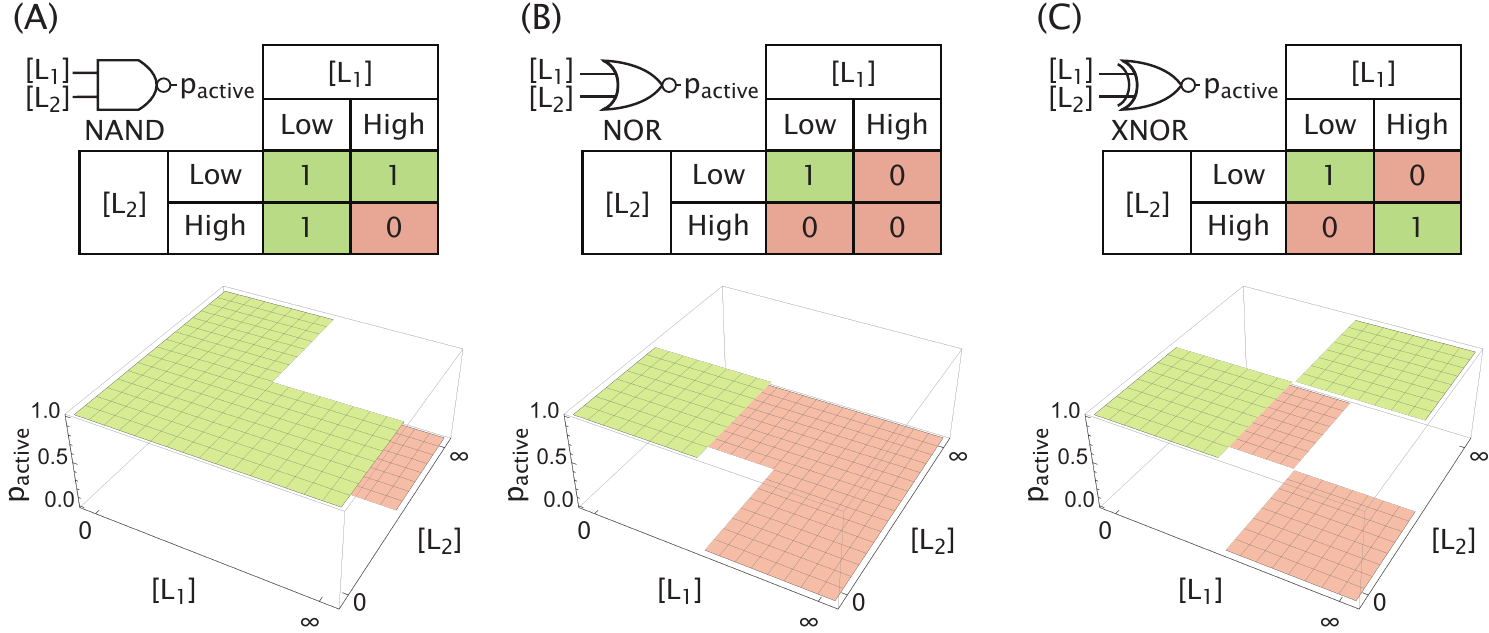}}
	\caption{\textbf{Additional logic gates as molecular responses.} The (A) NAND, (B) NOR, and
		(C) XNOR gates are the compliments of the AND, OR, and XOR gates, respectively, shown in Figure~\ref{fig:logic_gates}.} \label{fig:logic_gates_complements}
\end{figure}

In this section we derive the conditions necessary for an MWC molecule modulated by two ligands (with one binding site for each ligand) to exhibit the behavior of various logic gates shown in Figure~\ref{fig:logic_gates}. In addition to the three logic gates shown in Figure~\ref{fig:logic_gates}, we will also discuss the three complimentary gates NAND, NOR, and XNOR depicted in Figure~\ref{fig:logic_gates_complements}.

To simplify our notation, we define the value of $\text{p}_{\text{active}}$ from
eq~\ref{eq:mwc} in the following limits,
\begin{align} \label{eq:Concentrations_limits}
	\text{p}_{0,0} &= \text{p}_{\text{active}}([\text{L}_1] \rightarrow 0,[\text{L}_2] \rightarrow 0) = \frac{1}{1+ \text{e}^{-\text{\textbeta} \Delta \text{\textepsilon}_{\text{AI}}}}, \\
	\text{p}_{\infty,0} &= \text{p}_{\text{active}}([\text{L}_1] \rightarrow \infty,[\text{L}_2] \rightarrow 0) = \frac{1}{1+ \gamma_1 \text{e}^{-\text{\textbeta} \Delta \text{\textepsilon}_{\text{AI}}}}, \\
	\text{p}_{0,\infty} &= \text{p}_{\text{active}}([\text{L}_1] \rightarrow 0,[\text{L}_2] \rightarrow \infty) = \frac{1}{1+\gamma_2  \text{e}^{-\text{\textbeta} \Delta \text{\textepsilon}_{\text{AI}}}}, \\
	\text{p}_{\infty,\infty} &= \text{p}_{\text{active}}([\text{L}_1] \rightarrow \infty,[\text{L}_2] \rightarrow \infty) = \frac{1}{1+\gamma_1 \gamma_2 \text{e}^{-\text{\textbeta} \Delta \text{\textepsilon}_{\text{AI}}}},
\end{align}
where $\gamma_i =
\frac{\text{K}_{\text{A},\text{i}}}{\text{K}_{\text{I},\text{i}}}$ is the ratio
of the dissociation constants between the $\text{i}^\text{th}$ ligand and the
protein in the active and inactive states. From the ideal logic gate behaviors
visualized in Figure~\ref{fig:logic_gates} and
Figure~\ref{fig:logic_gates_complements}, we can then deduce the desired
constraints that model parameters need to meet for an effective realization of
each gate.

\subsection{AND Gate} \label{AND_gate_derivation}

Starting from the AND gate, we require $\text{p}_{0,0} \approx 0$, $\text{p}_{0,\infty} \approx 0$, $\text{p}_{\infty,0} \approx 0$ and $\text{p}_{\infty,\infty} \approx 1$, which yields the following conditions:
\begin{align}
\label{eqSI:AND1}
\text{e}^{-\text{\textbeta} \Delta \text{\textepsilon}_{\text{AI}}} \gg 1, \\
\label{eqSI:AND2}
\gamma_1 \text{e}^{-\text{\textbeta} \Delta \text{\textepsilon}_{\text{AI}}} \gg 1, \\
\gamma_2 \text{e}^{-\text{\textbeta} \Delta \text{\textepsilon}_{\text{AI}}} \gg 1 , \\
\label{eqSI:AND4}
\gamma_1 \gamma_2 \text{e}^{-\text{\textbeta} \Delta \text{\textepsilon}_{\text{AI}}} \ll 1.
\end{align}
Combining eqs~\ref{eqSI:AND2}-\ref{eqSI:AND4}, we obtain the condition for an AND gate, namely,
\begin{equation}
\label{eqSI:ANDdE}
\frac{1}{\gamma_1}, \frac{1}{\gamma_2} \ll \text{e}^{-\text{\textbeta} \Delta \text{\textepsilon}_{\text{AI}}} \ll \frac{1}{\gamma_1 \gamma_2}.
\end{equation}
Note, that the outer inequalities imply
\begin{align} \label{eqSI:ANDgamma}
\gamma_1, \gamma_2 \ll 1,
\end{align}
meaning that both ligands bind more tightly to the protein in the active than the inactive state.

\subsection{OR Gate} \label{OR_gate_derivation}

For $\text{p}_{\text{active}}$ to represent an OR gate across ligand concentration space, 
it must satisfy $\text{p}_{0,0}
\approx 0$, $\text{p}_{0,\infty} \approx 1$, $\text{p}_{\infty,0} \approx 1$ and
$\text{p}_{\infty,\infty} \approx 1$. This requires that the parameters obey
\begin{align}
\label{eqSI:OR1}
\text{e}^{-\text{\textbeta} \Delta \text{\textepsilon}_{\text{AI}}} \gg 1, \\
\gamma_1 \text{e}^{-\text{\textbeta} \Delta \text{\textepsilon}_{\text{AI}}} \ll 1, \\
\label{eqSI:OR3}
\gamma_2 \text{e}^{-\text{\textbeta} \Delta \text{\textepsilon}_{\text{AI}}} \ll 1 , \\
\label{eqSI:OR4}
\gamma_1 \gamma_2 \text{e}^{-\text{\textbeta} \Delta \text{\textepsilon}_{\text{AI}}} \ll 1.
\end{align}
Combining eqs~\ref{eqSI:OR1}-\ref{eqSI:OR3}, we obtain a constraint on the free energy difference,
\begin{equation} \label{eqSI:ORdE}
1 \ll \text{e}^{-\text{\textbeta} \Delta \text{\textepsilon}_{\text{AI}}} \ll \frac{1}{\gamma_1}, \frac{1}{\gamma_2}.
\end{equation}
As with the AND gate, the outer inequalities imply that the ligands prefer binding to the protein in the active state,
\begin{equation} \label{eqSI:ORgamma}
\gamma_1, \gamma_2 \ll 1.
\end{equation}

\subsection{NAND and NOR Gates}
\label{appendix:NANDandNOR}
Because the NAND and NOR gates are the logical complements of AND and OR gates, 
respectively, the parameter constraints under which they are realized are the opposites
of those for AND and OR gates. Hence, the conditions for a NAND gate are given by
\begin{equation}
\frac{1}{\gamma_1 \gamma_2} \ll \text{e}^{-\text{\textbeta} \Delta \text{\textepsilon}_{\text{AI}}} \ll \frac{1}{\gamma_1}, \frac{1}{\gamma_2}
\end{equation}
while the conditions for NOR gates are 
\begin{equation}
\frac{1}{\gamma_1}, \frac{1}{\gamma_2} \ll \text{e}^{-\text{\textbeta} \Delta \text{\textepsilon}_{\text{AI}}} \ll 1.
\end{equation}
We note that in both cases, the outer inequalities imply that both ligands bind
more tightly to the protein in the inactive state than in the active state,
$\gamma_1, \gamma_2 \gg 1$. 

The symmetry between AND/OR and NAND/NOR gates also implies a simple relation between their
quality metrics, namely, $\text{Q}_{\text{AND/OR}} \left( \gamma_1, \gamma_2, \Delta
\text{\textepsilon}_{\text{AI}} \right) = \text{Q}_{\text{NAND/NOR}} \left(
\frac{1}{\gamma_1}, \frac{1}{\gamma_2}, -\Delta \text{\textepsilon}_{\text{AI}}
\right)$. Here we provide a proof for the AND gate and invite the reader to do the same for the OR gate.
From eq~\ref{eq:quality}, the quality metrics for the AND and NAND gates can be written as
\begin{align}
\text{Q}_{\text{AND}} (\gamma_1, \gamma_2, \omega) &= (1-\text{p}_{0,0})(1-\text{p}_{\infty,0})(1-\text{p}_{0,\infty}) \text{p}_{\infty,\infty} \nonumber\\
&= \left( 1 - \frac{1}{1+\omega}\right) \left( 1 - \frac{1}{1+\gamma_1 \omega} \right) \left( 1 - \frac{1}{1+\gamma_2 \omega} \right) \left( \frac{1}{1+\gamma_1\gamma_2 \omega}  \right) \nonumber\\
&= \frac{\gamma_1 \gamma_2 \omega^3}{(1+\omega)(1+\gamma_1 \omega) (1+\gamma_2 \omega) (1+\gamma_1 \gamma_2 \omega)}, \\
\label{eqn:QNAND}
\text{Q}_{\text{NAND}} (\gamma_1, \gamma_2, \omega) &= \text{p}_{0,0} \text{p}_{\infty,0} \text{p}_{0,\infty} (1-\text{p}_{\infty,\infty}) \nonumber\\
&= \left( \frac{1}{1+\omega} \right) \left( \frac{1}{1+\gamma_1 \omega}  \right) \left( \frac{1}{1+\gamma_2 \omega} \right)  \left(1 - \frac{1}{1+\gamma_1 \gamma_2 \omega}  \right) \nonumber\\
&= \frac{\gamma_1 \gamma_2 \omega}{(1+\omega)(1+\gamma_1 \omega) (1+\gamma_2 \omega) (1+\gamma_1 \gamma_2 \omega)},
\end{align}
where we introduced $\omega = \text{e}^{-\text{\textbeta} \Delta \text{\textepsilon}_{\text{AI}}}$. 
Substituting $\gamma_1 \rightarrow \gamma_1^{-1}, \gamma_2 \rightarrow \gamma_2^{-1}, \omega \rightarrow \omega^{-1}$ (equivalent to $\Delta
\text{\textepsilon}_{\text{AI}} \rightarrow -\Delta \text{\textepsilon}_{\text{AI}}$) in eq~\ref{eqn:QNAND}, we obtain
\begin{align}
\text{Q}_{\text{NAND}} ( \gamma_1^{-1}, \gamma_2^{-1}, \omega^{-1}) 
&= \frac{\gamma_1^{-1} \gamma_2^{-1} \omega^{-1}}{(1+\omega^{-1})(1+\gamma_1^{-1} \omega^{-1}) (1+\gamma_2^{-1} \omega^{-1}) (1+\gamma_1^{-1} \gamma_2^{-1} \omega^{-1})} \times \frac{\gamma_1^2 \gamma_2^2 \omega^4}{\gamma_1^2 \gamma_2^2 \omega^4} \nonumber\\
&= \frac{\gamma_1 \gamma_2 \omega^3 }{(1+\omega)(1+\gamma_1 \omega) (1+\gamma_2 \omega) (1+\gamma_1 \gamma_2 \omega)} \nonumber\\
&\equiv\text{Q}_{\text{AND}} (\gamma_1, \gamma_2, \omega).
\end{align}

\subsection{XOR and XNOR Gates}
\label{Appendix_XOR}

Here, we show that the XOR gate (and by symmetry the XNOR gate) are not
achievable with the form of $\text{p}_\text{active}$ given in eq~\ref{eq:mwc}. An
XOR gate satisfies $\text{p}_{0,0} \approx 0$, $\text{p}_{0,\infty} \approx 1$,
$\text{p}_{\infty,0} \approx 1$ and $\text{p}_{\infty,\infty} \approx 0$ which
necessitates the parameter conditions
\begin{align}
\text{e}^{-\text{\textbeta} \Delta \text{\textepsilon}_{\text{AI}}} \gg 1, \label{eqSI:XOR1} \\
\gamma_1 \text{e}^{-\text{\textbeta} \Delta \text{\textepsilon}_{\text{AI}}} \ll 1, \label{eqSI:XOR2} \\
\gamma_2 \text{e}^{-\text{\textbeta} \Delta \text{\textepsilon}_{\text{AI}}} \ll 1, \label{eqSI:XOR3} \\
\gamma_1 \gamma_2 \text{e}^{-\text{\textbeta} \Delta \text{\textepsilon}_{\text{AI}}} \gg 1. \label{eqSI:XOR4}
\end{align}
However, these conditions cannot all be satisfied, as the left-hand side of eq~\ref{eqSI:XOR4} can be written in terms of the left-hand sides of eqs~\ref{eqSI:XOR1}-\ref{eqSI:XOR3},  
\begin{equation} \label{eqSI:XOR4_rewrite}
\gamma_1 \gamma_2 \text{e}^{-\text{\textbeta} \Delta \text{\textepsilon}_{\text{AI}}} = \frac{\left( \gamma_1
	\text{e}^{-\text{\textbeta} \Delta \text{\textepsilon}_{\text{AI}}} \right) \left(
	\gamma_2 \text{e}^{-\text{\textbeta} \Delta \text{\textepsilon}_{\text{AI}}}
	\right)}{\text{e}^{-\text{\textbeta} \Delta \text{\textepsilon}_{\text{AI}}}} \ll 1,
\end{equation}
contradicting eq~\ref{eqSI:XOR4}.

\begin{figure}[!ht]
	\centerline{\includegraphics{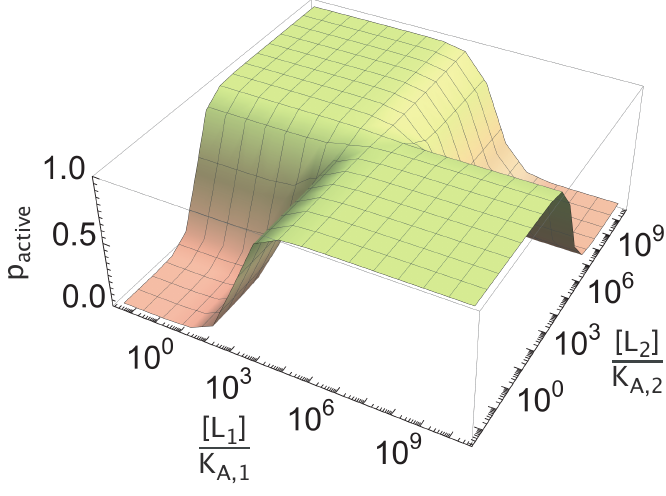}}	
	\caption{\textbf{An XOR gate can be achieved by adding cooperativity.} The
	activity profile defined in eq~\ref{eq:mwcCooperativity} for the parameter values from
	Figure~\ref{fig:MWC_simple}B, along with the cooperativity energies
	$\text{\textepsilon}_{\text{A,coop}} = 15\,\text{k}_\text{B} \text{T}$ and
	$\text{\textepsilon}_{\text{I,coop}} = 0$, give rise to an XOR response.}
\label{fig:XORgate}
\end{figure}

The XOR gate could be realized if an explicit cooperativity energy
$\text{\textepsilon}_{\text{A,coop}}$ is added when both ligands are bound in
the active state and $\text{\textepsilon}_{\text{I,coop}}$ when both are bound
in the inactive state. These cooperative interactions modify eq~\ref{eq:mwc}
to the form
\begin{equation} \label{eq:mwcCooperativity}
\text{p}_{\text{active}}\left( [\text{L}_1], [\text{L}_2] \right) = \frac{1 + \frac{[\text{L}_1]}{\text{K}_{\text{A},1}} + \frac{[\text{L}_2]}{\text{K}_{\text{A},2}} + \frac{[\text{L}_1]}{\text{K}_{\text{A},1}} \frac{[\text{L}_2]}{\text{K}_{\text{A},2}} \text{e}^{-\text{\textbeta} \text{\textepsilon}_{\text{A,coop}}}}{1 + \frac{[\text{L}_1]}{\text{K}_{\text{A},1}} + \frac{[\text{L}_2]}{\text{K}_{\text{A},2}} + \frac{[\text{L}_1]}{\text{K}_{\text{A},1}} \frac{[\text{L}_2]}{\text{K}_{\text{A},2}} \text{e}^{-\text{\textbeta} \text{\textepsilon}_{\text{A,coop}}} + \text{e}^{-\text{\textbeta} \Delta \text{\textepsilon}_{\text{AI}}} \left( 1 + \frac{[\text{L}_1]}{\text{K}_{\text{I},1}} + \frac{[\text{L}_2]}{\text{K}_{\text{I},2}} + \frac{[\text{L}_1]}{\text{K}_{\text{I},1}} \frac{[\text{L}_2]}{\text{K}_{\text{I},2}} \text{e}^{-\text{\textbeta} \text{\textepsilon}_{\text{I,coop}}} \right)}.
\end{equation}
Figure~\ref{fig:XORgate} demonstrates that the same parameter values from
Figure~\ref{fig:MWC_simple}B together with the (unfavorable) cooperativity
energy $\text{\textepsilon}_{\text{A,coop}} = 15\,\text{k}_\text{B} \text{T}$ and
$\text{\textepsilon}_{\text{I,coop}} = 0$ can create an XOR gate.

\pagebreak
\section{The General Two-Ligand Response: Transitioning Between OFF and ON States} \label{Appendix_general_response}

In the preceding section, we have been solely concerned with the
behavior of the MWC molecule in the limits of ligand concentration
($[\text{L}_\text{i}]=0$ and $[\text{L}_\text{i}] \to \infty$), and have ignored
the details about the transition from ON to OFF (e.g., its shape and
steepness) and also the possibility of $\text{p}_{\text{active}} \neq 0$ or $1$.
In this section, we examine and derive in greater detail some of the additional response
behaviors that are possible for an MWC molecule regulated with $\text{N}=2$ ligands when
the locations of transitions between limit responses are taken into account.

To examine the transitions between $\text{p}_{\text{active}}$ levels, we
 derive expressions for the concentrations at which transitions are at their midpoint. Since
$\text{p}_{\text{active}}$ is a function of two different ligand concentrations,
[$\text{L}_1$] and [$\text{L}_2$], we define two different midpoint
concentrations of ligand $\text{L}_{\text{i}}$: one in the absence of ligand
$\text{L}_{\text{j}}$, $[\text{L}_{\text{i}}^*]_{[\text{L}_{\text{j}}] \to 0}$, and another when  $\text{L}_{\text{j}}$ is
saturating, $[\text{L}_{\text{i}}^*]_{[\text{L}_{\text{j}}] \to \infty}$. In particular, $[\text{L}_{\text{i}}^*]_{[\text{L}_{\text{j}}] \to 0}$ is defined such that
\begin{equation}\label{eqSI:transition_definition_1_absent}
\text{p}_{\text{active}} \left( [\text{L}_{\text{i}}^*]_{[\text{L}_{\text{j}}] \to 0} , 
[\text{L}_{\text{j}}]=0 \right) = 
\frac{
\text{p}_{\text{active}} \left( [\text{L}_{\text{i}}]=0, 
[\text{L}_{\text{j}}]=0 \right)
+
\text{p}_{\text{active}} \left( [\text{L}_{\text{i}}] \to \infty, 
[\text{L}_{\text{j}}]=0
\right)}
{2},
\end{equation}
i.e., the concentration of ligand i where $\text{p}_{\text{active}}$ is equal to
the mean of the two $\text{p}_{\text{active}}$ limit values being transitioned
between. If we evaluate the left hand side of 
eq~\ref{eqSI:transition_definition_1_absent} with i = 1 and j = 2 using
eq~\ref{eq:mwc}, and the right hand side using the limits from Figure \ref{fig:MWC_simple}(A), we obtain
\begin{align}\label{eqSI:transition_equation_1_absent}
 \frac{\left( 1 + \frac{[\text{L}_1^*]_{[\text{L}_{\text{2}}] \to 0}}{\text{K}_{\text{A},1}} \right)}
 {\left( 1 + \frac{[\text{L}_1^*]_{[\text{L}_{\text{2}}] \to 0}}{\text{K}_{\text{A},1}} \right) + 
 \text{e}^{-\text{\textbeta} \Delta \text{\textepsilon}_{\text{AI}}} \left( 1 + \frac{[\text{L}_1^*]_{[\text{L}_{\text{2}}] \to 0}}{\text{K}_{\text{I},1}} \right)} = 
 \frac{1}{2}\left(\frac{1}{1+  \text{e}^{-\text{\textbeta} \Delta \text{\textepsilon}_{\text{AI}}}} +  
\frac{1}{1+  \gamma_1 \, \text{e}^{-\text{\textbeta} \Delta \text{\textepsilon}_{\text{AI}}}} \right).
\end{align}
Introducing $\gamma_{\text{1}} = \text{K}_{\text{A},1}/\text{K}_{\text{I},1}$, we can solve for $[\text{L}_1^*]_{[\text{L}_{\text{2}}] \to 0}$ to find
\begin{equation}\label{eqSI:transition_concentration_1_absent}
\frac{[\text{L}_{\text{1}}^*]_{{[\text{L}_{\text{2}}] \to 0}}}{\text{K}_{\text{A},\text{1}}} = \frac{1+\text{e}^{-\text{\textbeta} \Delta \text{\textepsilon}_{\text{AI}}}}{1+\gamma_1 \, \text{e}^{-\text{\textbeta} \Delta \text{\textepsilon}_{\text{AI}}}}.
\end{equation}
Eq~\ref{eqSI:transition_concentration_1_absent} can be rewritten for
$[\text{L}_{\text{2}}^*]_{[\text{L}_{\text{1}}] \to 0}$ by merely
interchanging all ligand and parameter indices, i.e., 1 $\leftrightarrow$ 2.

The midpoint concentration when one ligand is saturating can be derived
similarly. Specifically, to find an expression for
$[\text{L}_{\text{i}}^*]_{[\text{L}_{\text{j}}] \to \infty}$ we can
re-write~\ref{eqSI:transition_definition_1_absent} using eq~\ref{eq:mwc} in the
case that [$\text{L}_{\text{j}}] \to \infty$ with i = 1 and j = 2, resulting in
\begin{align}\label{eqSI:transition_equation_1_saturating}
 \frac{\left( 1 + \frac{[\text{L}_1^*]_{[\text{L}_{\text{2}}] \to \infty}}{\text{K}_{\text{A},1}} \right)}
 {\left( 1 + \frac{[\text{L}_1^*]_{[\text{L}_{\text{2}}] \to \infty}}{\text{K}_{\text{A},1}} \right) + 
 \gamma_2 \text{e}^{-\text{\textbeta} \Delta \text{\textepsilon}_{\text{AI}}} \left( 1 + \frac{[\text{L}_1^*]_{[\text{L}_{\text{2}}] \to \infty}}{\text{K}_{\text{I},1}} \right)} = 
  \frac{1}{2}\left(\frac{1}{1+  \gamma_2 \, \text{e}^{-\text{\textbeta} \Delta \text{\textepsilon}_{\text{AI}}}} +  
\frac{1}{1+  \gamma_1 \gamma_2\, \text{e}^{-\text{\textbeta} \Delta \text{\textepsilon}_{\text{AI}}}} \right).
 \end{align}
Eq~\ref{eqSI:transition_equation_1_saturating} can be solved for $[\text{L}_{\text{1}}^*]_{[\text{L}_{\text{2}}] \to \infty}$ to produce,
\begin{equation}\label{eqSI:transition_concentration_1_saturating}
\frac{[\text{L}_{\text{1}}^*]_{{[\text{L}_{\text{2}}] \to \infty}}}{\text{K}_{\text{A},\text{1}}} = \frac{1+ \gamma_2 \, \text{e}^{-\text{\textbeta} \Delta \text{\textepsilon}_{\text{AI}}}}{1+\gamma_1 \gamma_2\, \text{e}^{-\text{\textbeta} \Delta \text{\textepsilon}_{\text{AI}}}}.
\end{equation}
Again, the symmetric expression for $[\text{L}_{\text{2}}^*]_{{[\text{L}_{\text{1}}] \to \infty}}$ is found by 
swapping ligand and parameter indices, 1$\leftrightarrow$2.

Using this approach to define concentration transition zones can be used to produce additional MWC behaviors, including the ratiometric response in the BMP pathway recently analyzed by Antebi {\it et al.}~\cite{Antebi2017}, which was briefly discussed earlier. Specifically, this response can be approximated by choosing parameter values that satisfy two desired limits, $\text{p}_{\infty,0} \approx 0$ ($\gamma_1 \, \text{e}^{-\text{\textbeta} \Delta \text{\textepsilon}_{\text{AI}}} \gg 1$) and $\text{p}_{0,\infty}\approx 1$ ($\gamma_2 \, \text{e}^{-\text{\textbeta} \Delta \text{\textepsilon}_{\text{AI}}} \ll 1$), as well as produce a large transition region sensitive to both ligands, i.e., the ratio in eq~\ref{eq:transition_ratio}, $\frac{[\text{L}_{\text{i}}^*]_{{[\text{L}_{\text{j}}] \to \infty}}}{[\text{L}_{\text{i}}^*]_{{[\text{L}_{\text{j}}] \to 0}}}$ is far from 1.
One way to satisfy these conditions is to set $\text{K}_{\text{I},\text{2}} \gg \text{K}_{\text{A},\text{1}} = \text{K}_{\text{A},\text{2}} \gg \text{K}_{\text{I},\text{1}}$ and $\Delta
\text{\textepsilon}_{\text{AI}} = 0$ in eq~\ref{eq:mwc}. 
Notice that with these parameter choices and provided the ligand concentrations satisfy
\begin{align} \label{eqSI:ratiometricCriteria}
\frac{[\text{L}_1]}{\text{K}_{\text{A},1}}, \frac{[\text{L}_2]}{\text{K}_{\text{I},2}} &\ll 1, \nonumber \\
\frac{[\text{L}_1]}{\text{K}_{\text{I},1}}, \frac{[\text{L}_2]}{\text{K}_{\text{A},2}} &\gg 1,
\end{align}
the probability that the protein is active reduces to
\begin{equation}
\text{p}_{\text{active}}\left( [\text{L}_1], [\text{L}_2] \right) \approx \frac{\frac{[\text{L}_2]}{\text{K}_{\text{A},2}}}{\frac{[\text{L}_2]}{\text{K}_{\text{A},2}} + \frac{[\text{L}_1]}{\text{K}_{\text{I},1}}}. 
\end{equation}
Hence, only the ratio of $[\text{L}_1]$ and $[\text{L}_2]$ matters, as shown in
Figure~\ref{fig:general_two_ligand_response}B where eq~\ref{eqSI:ratiometricCriteria} is
satisfied provided that $10^{-4} \lesssim
\frac{[\text{L}_1]}{\text{K}_{\text{A},1}} \lesssim 10^0 \lesssim
\frac{[\text{L}_2]}{\text{K}_{\text{A},2}} \lesssim 10^4$.

Additionally, we consider the remaining three types of input-output computations shown by Antebi
\textit{et al.}~to exist in the BMP pathway which they called the additive,
imbalance, and balance responses \cite{Antebi2017}. The additive response
(which responds more to larger input concentrations) is an OR gate which we
showed is possible in Figure~\ref{fig:MWC_simple}B. The imbalance response (which
responds maximally to extreme ratios of the two input ligands) is similar to an XOR behavior
which, as discussed in Appendix~\ref{Appendix_XOR}, is only achievable with an
explicit cooperativity energy.

\begin{figure}[!ht]
	\centerline{\includegraphics{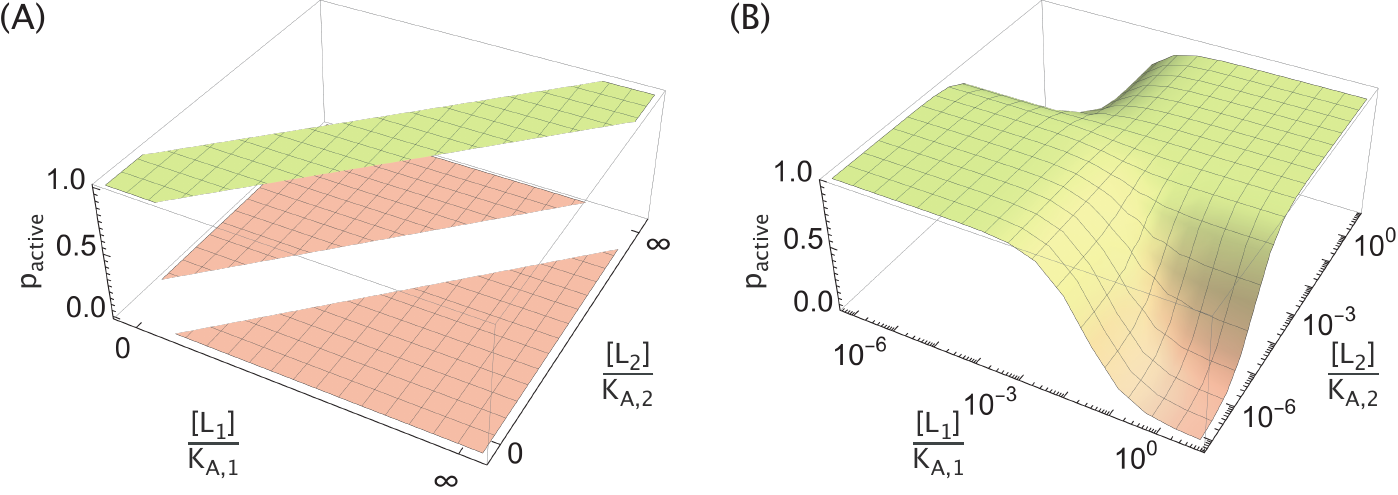}}
	\caption{\textbf{Balance response behavior approximated by the MWC model.} (A)
	The ideal balance response from the BMP pathway and (B) the closest behavior that an MWC molecule can
	exhibit using the complementary parameters from Figure~\ref{fig:XORgate}
	($\text{K}_{\text{A},\text{i}} = 1.5 \times 10^{-4}\,\text{M}$,
	$\text{K}_{\text{I},\text{i}} = 2.5 \times 10^{-8}\,\text{M}$, $\Delta
	\text{\textepsilon}_{\text{AI}} = 5\,\text{k}_{\text{B}} \text{T}$,
	$\text{\textepsilon}_{\text{A,coop}} = -15\,\text{k}_\text{B} \text{T}$ and
	$\text{\textepsilon}_{\text{I,coop}} = 0$).}
\label{fig:OtherGates}
\end{figure}

The balance response is defined as
\begin{equation} \label{eq:balance}
	\text{p}_{\text{active}}^{\text{balance}} = 
	\begin{cases}
	1 & [\text{L}_1] \approx [\text{L}_2] \\
	0 & [\text{L}_1] \not\approx [\text{L}_2]
	\end{cases}
\end{equation}
so that the protein is only ON when both ligands are present in the same amount
as shown in Figure~\ref{fig:OtherGates}A. Such behavior is not possible within
the MWC model because starting from any point $\overline{[\text{L}_1]} =
\overline{[\text{L}_2]}$, $\text{p}_{\text{active}}$ in eq~\ref{eq:mwc} must
either monotonically increase or monotonically decrease with $[\text{L}_1]$
(depending on $\gamma_1$), whereas eq~\ref{eq:balance} requires that
$\text{p}_{\text{active}}$ must decrease for both $[\text{L}_1] >
\overline{[\text{L}_1]}$ and $[\text{L}_1] < \overline{[\text{L}_1]}$ (with
similar contradictory statements for $[\text{L}_2]$). The closest behavior
achievable by the MWC model is to zoom into the transition region of an XNOR gate
as shown in Figure~\ref{fig:OtherGates}B. As we zoom out of the concentration
ranges shown, the four square regions of the plot will continue to expand as
squares and the behavior will no longer approximate the ideal balance response.

\pagebreak
\section{Logic Switching by Tuning the Number of Ligand Binding Sites}
\label{Appendix_increased_binding_sites}

In this section, we show how an MWC molecule whose activity is given by
eq~\ref{eq:mwc_generalize} can switch between exhibiting AND$\leftrightarrow$OR
or NAND$\leftrightarrow$NOR behaviors by tuning the number of binding sites. 
To begin, we define the probability $\text{p}_\text{active}$ that the molecule is
active in the case when the  $\text{i}^{\text{th}}$ ligand has $\text{n}_{\text{i}}$ binding sites, namely, 
\begin{align}
\text{p}_{0,0} &= \text{p}_{\text{active}}([\text{L}_1] \rightarrow 0,[\text{L}_2] \rightarrow 0) = \frac{1}{1+ \text{e}^{-\text{\textbeta} \Delta \text{\textepsilon}_{\text{AI}}}}, \\
\text{p}_{\infty,0} &= \text{p}_{\text{active}}([\text{L}_1] \rightarrow \infty,[\text{L}_2] \rightarrow 0) = \frac{1}{1+ \gamma_1^{\text{n}_1} \text{e}^{-\text{\textbeta} \Delta \text{\textepsilon}_{\text{AI}}}}, \\
\text{p}_{0,\infty} &= \text{p}_{\text{active}}([\text{L}_1] \rightarrow 0,[\text{L}_2] \rightarrow \infty) = \frac{1}{1+\gamma_2^{\text{n}_2}  \text{e}^{-\text{\textbeta} \Delta \text{\textepsilon}_{\text{AI}}}}, \\
\text{p}_{\infty,\infty} &= \text{p}_{\text{active}}([\text{L}_1] \rightarrow \infty,[\text{L}_2] \rightarrow \infty) = \frac{1}{1+ \gamma_1^{\text{n}_1} \gamma_2^{\text{n}_2} \text{e}^{-\text{\textbeta} \Delta \text{\textepsilon}_{\text{AI}}}}.
\end{align}
Note that the only effect of having an arbitrary number of
ligand binding sites (as opposed to
$\text{n}_\text{i} = 1$ as in Appendix~\ref{Appendix_condition_derivations}) is
that the ratio of dissociation constants always appears raised to the number of
binding sites, $\gamma_i^{\text{n}_\text{i}}$. Hence, the parameter conditions 
derived for AND and OR behaviors for $\text{n}_{\text{i}} = 1$ can be 
used in the case of general $\text{n}_{\text{i}}$ by substituting $\gamma_\text{i} \to
\gamma_\text{i}^{\text{n}_\text{i}}$.

Now, suppose a molecule with $\text{N}=2$ ligands and with $\text{n}^\prime_1$ and $\text{n}^\prime_2$
binding sites for ligands 1 and 2 represents an AND gate, while this same molecule
with $\text{n}_1$ and $\text{n}_2$ binding sites serves as an OR gate, as in
Figure~\ref{fig:mwc_generalizations}B with
$\text{n}^\prime_1=\text{n}^\prime_2=1$ and $\text{n}_1=\text{n}_2=4$.
 From Figure~\ref{fig:MWC_simple}B, the conditions in the former case (AND gate) are
\begin{equation}
\frac{1}{\gamma_1^{\text{n}^\prime_1}}, \frac{1}{\gamma_2^{\text{n}^\prime_2}} \ll \text{e}^{-\text{\textbeta} \Delta \text{\textepsilon}_{\text{AI}}} \ll \frac{1}{\gamma_1^{\text{n}^\prime_1} \gamma_2^{\text{n}^\prime_2}},
\end{equation}
while the conditions in the latter case (OR gate) are
\begin{equation}
1 \ll \text{e}^{-\text{\textbeta} \Delta \text{\textepsilon}_{\text{AI}}} \ll \frac{1}{\gamma_1^{\text{n}_1}}, \frac{1}{\gamma_2^{\text{n}_2}}.
\end{equation}
Combining these conditions, we find that the requirements for the AND$\leftrightarrow$OR switching are given by
\begin{equation} \label{eq:classSwitchingRequirement2}
\frac{1}{\gamma_1^{\text{n}^\prime_1}}, \frac{1}{\gamma_2^{\text{n}^\prime_2}} \ll \text{e}^{-\text{\textbeta} \Delta \text{\textepsilon}_{\text{AI}}} \ll \frac{1}{\gamma_1^{\text{n}_1}}, \frac{1}{\gamma_2^{\text{n}_2}}, \frac{1}{\gamma_1^{\text{n}^\prime_1} \gamma_2^{\text{n}^\prime_2}},
\end{equation}
where we have used the fact that the outer inequalities imply
$\gamma_1^{\text{n}^\prime_1},\gamma_2^{\text{n}^\prime_2} \ll 1$ (so that $1 \ll \frac{1}{\gamma_1^{\text{n}^\prime_1}}, \frac{1}{\gamma_2^{\text{n}^\prime_2}}$). In the limit
$\text{n}^\prime_1=\text{n}^\prime_2=1$, eq~\ref{eq:classSwitchingRequirement2}
reduces to the condition shown in Figure~\ref{fig:mwc_generalizations}A.

Lastly, we note that since NAND is the complement of AND while NOR is the
complement of OR, the class switching requirements in
\ref{eq:classSwitchingRequirement2} become the
requirements to change from NAND$\leftrightarrow$NOR behavior when
$\gamma_\text{i} \to \frac{1}{\gamma_\text{i}}$ and $\Delta
\text{\textepsilon}_{\text{AI}} \to -\Delta \text{\textepsilon}_{\text{AI}}$.

\pagebreak

\section{Combinatorial Control with Three Regulatory Ligands}
\label{Appendx_three_ligands}

In this section, we first present the methodology used to identify the functionally
unique and MWC-compatible 3-ligand logic gates.
We then use the full list of admissible gates to find all possible logic switches
that can be induced by increasing the concentration of a third ligand.
We finish the section by deriving the parameter conditions required
for achieving the logic switches {AND$\to$OR} and {AND$\to$YES$_1$}
shown in Figure \ref{fig:exampleSwitches}D.

\subsection{Functionally Unique MWC Gates}
\label{AppendixC_UniqueGates}

To identify the set of functionally unique MWC gates, we first iterate over the 256 possible 
responses and eliminate those redundant ones that can be obtained by shuffling the ligand labels 
of already selected gates. The Python implementation of this procedure that leaves 80 
functionally unique gates can be found in the supplementary Jupyter Notebook 1.

\begin{figure}[!ht]
	\centerline{\includegraphics{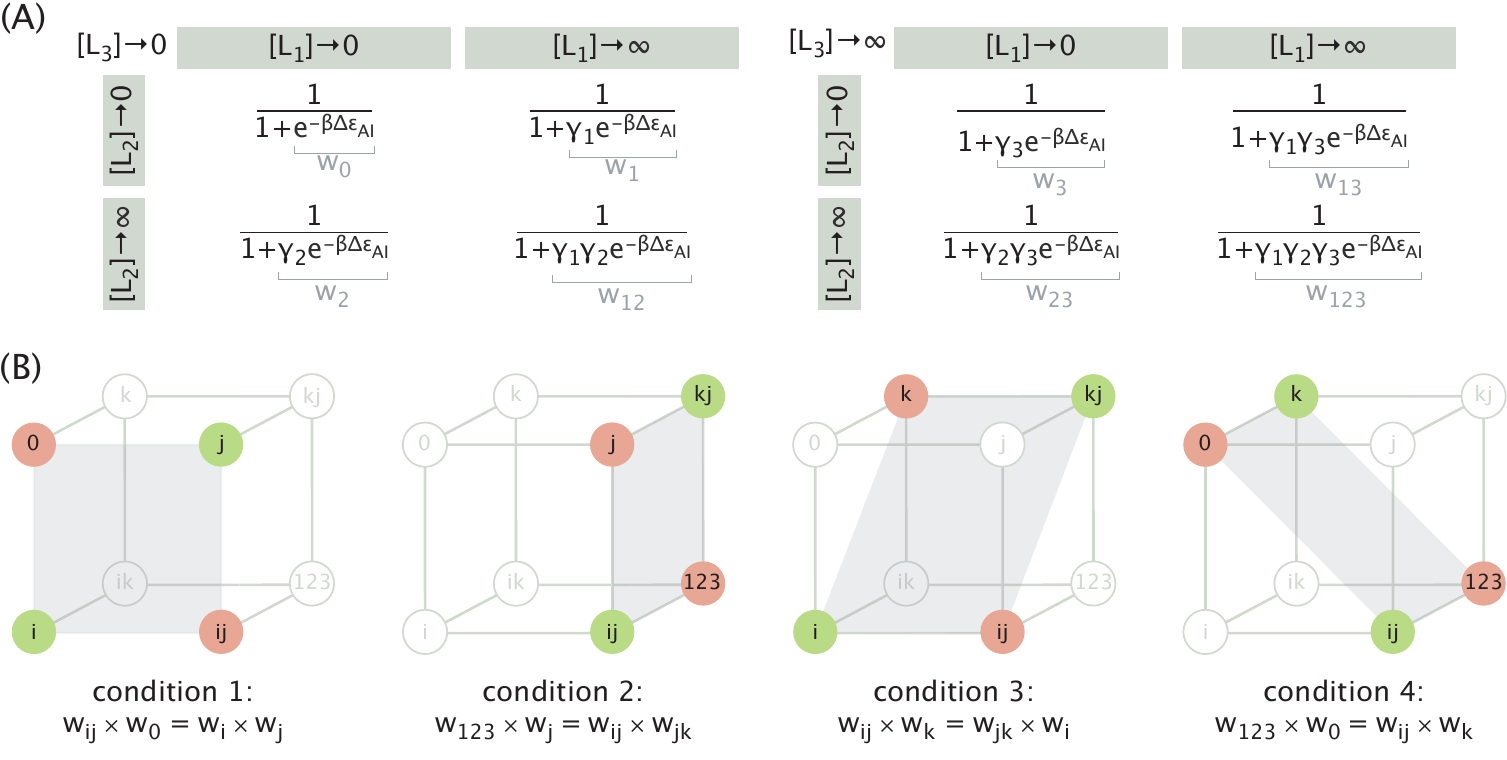}}
	\caption{\textbf{Three-ligand logic gates that are incompatible with the MWC framework.}
	(A) Probability that the protein is active in the 8
		different ligand concentration limits. The total weight of the inactive states
		relative to the active states is indicated in gray for all limits.
	(B) Cubic diagrams of logic
		responses that are incompatible with the MWC framework, along with the
		constraint equations used to obtain them. The limits relevant to the
		constraint conditions are shown in color, and a transparent gray plane containing these
		relevant limits is added for clarity. In all four diagrams $1 \le
		\text{i},\text{j},\text{k} \le 3$.}
	\label{figS:MWC_constraints}
\end{figure}

Having singled out the functionally unique responses, we proceed to identify
those that are admissible in the MWC framework. To that end, we first write the
analytic forms for the probability of the protein being active
($\text{p}_{\text{active}}$) at eight different ligand concentration limits (Figure
\ref{figS:MWC_constraints}A). Since the functional form in all cases is
 $\text{p}_{\text{active}} = (1+\text{w}_{\text{I/A}})^{-1}$, where
$\text{w}_{\text{I/A}}$ is the total weight of the inactive states divided by
	the total weight of the active states in the appropriate limit (as seen in
	Figure~\ref{fig:MWC_simple}A), a Boolean response ($\text{p}_{\text{active}}
\approx 0 \,\, \text{or} \, 1$) can only be achieved when $\text{w}_{\text{I/A}}
\gg 1$ or $\text{w}_{\text{I/A}} \ll 1$, respectively. Hence, the values of
$\text{w}_{\text{I/A}}$ at the eight different limits of ligand concentration will
determine the full logic response of the protein.

Note that since cooperative interactions between ligands are absent in the MWC
framework, the eight different $\text{w}_{\text{I/A}}$ expressions depend on only
four independent MWC parameters, namely, $\{ \Delta
\text{\textepsilon}_{\text{AI}}, \gamma_1, \gamma_2, \gamma_3 \}$.
Therefore, only four of the eight limiting $\text{w}_{\text{I/A}}$ values can be independently tuned,
and any $\text{w}_{\text{I/A}}$ limit can be expressed as a function of four different and
independent $\text{w}_{\text{I/A}}$ limits,
resulting in a constraint condition. Since each $\text{w}_{\text{I/A}}$ is a product of some
$\gamma_\text{i}$'s and $\text{e}^{-\text{\textbeta} \Delta \text{\textepsilon}_{\text{AI}}}$
(Figure \ref{figS:MWC_constraints}A), we look for constraint conditions
that have a multiplicative form, namely,
\begin{align}
\text{w}_{\text{s}^*} = \prod_{\text{i} = 1}^4 \text{w}_{\text{s}_\text{n}}^{\alpha_{\text{n}}},
\end{align}
where $\text{w}_{\text{s}^*}$ is the target limit, $\text{s}_\text{n} \ne  \text{s}^* (1 \le \text{n} \le 4)$ are the labels of four different limits and 
$\alpha_\text{n}$ are real coefficients.
Searching over all conditions of such form (see the supplementary Jupyter Notebook 2 for details),
we identify a total of eight functionally unique constraints,
\begin{align}
\text{w}_{\text{ij}} \times \text{w}_{\text{0}} &= \text{w}_{\text{i}} \times \text{w}_{\text{j}},  \label{eqnS:cond1}\\
\text{w}_{\text{123}} \times \text{w}_{\text{j}} &= \text{w}_{\text{ij}} \times \text{w}_{\text{jk}}, \\
\text{w}_{\text{ij}} \times \text{w}_{\text{k}} &= \text{w}_{\text{ik}} \times \text{w}_{\text{j}},  \\
\text{w}_{\text{123}} \times \text{w}_{\text{0}} &= \text{w}_{\text{ij}} \times \text{w}_{\text{k}}, \label{eqnS:cond4}\\
\text{w}_{\text{ij}} \times \text{w}_{\text{k}}^2 &= \text{w}_{\text{0}} \times \text{w}_{\text{ik}} \times \text{w}_{\text{jk}}, \\
\text{w}_{\text{123}} \times \text{w}_{\text{0}}^2 &= \text{w}_{\text{1}} \times \text{w}_{\text{2}} \times \text{w}_{\text{3}}, \\
\text{w}_{\text{123}}^2 \times \text{w}_{\text{0}} &= \text{w}_{\text{12}} \times \text{w}_{\text{13}} \times \text{w}_{\text{23}}, \\
\text{w}_{\text{123}} \times \text{w}_{\text{i}} \times \text{w}_{\text{j}} &= \text{w}_{\text{ij}}^2 \times \text{w}_{\text{k}},
\end{align}
where $1 \le \text{i, j, k} \le 3$.

Further searching for a minimum set of constraints that can account for all gates incompatible with the
MWC framework, we identify the constraints in eqs \ref{eqnS:cond1}-\ref{eqnS:cond4} as the necessary and sufficient ones (see the supplementary 
Jupyter Notebook 2). Graphical representations of these four constraints on a cubic diagram are shown in Figure \ref{figS:MWC_constraints}B.
Note that these conditions are all of the form
\begin{align}
\text{w}_{\text{s}_1} \text{w}_{\text{s}_2} = \text{w}_{\text{s}_3} \text{w}_{\text{s}_4},
\end{align}
where $\text{s}_{\text{i}}$ are labels corresponding to different ligand
concentration limits. Logic responses where $\text{w}_{\text{s}_1},
\text{w}_{\text{s}_2} \ll 1 \, (\gg 1)$ while $\text{w}_{\text{s}_3},
\text{w}_{\text{s}_4} \gg 1 \, (\ll 1)$ cannot be achieved, since they
contradict the constraint condition. Conditions 1 and 2 in Figure
\ref{figS:MWC_constraints}B, for example, demonstrate that XOR and XNOR gates
cannot be realized by any two ligands in the absence (condition 1) or presence
(condition 2) of a third ligand - a result expected from the 2-ligand analysis
done earlier. On the other hand, conditions 3 and 4 are specific to the
3-ligand response.

Checking the 80 functionally unique gates against the four constraints in 
Figure \ref{figS:MWC_constraints}B, we obtain a set of 34 functionally unique
and MWC-compatible gates, 17 of which are shown in Figure \ref{figS:MWC_all}A
while the other half are their logical complements (i.e. ON$\leftrightarrow$OFF swapping
is performed for each of the cube elements).

\newpage
\subsection{Logic Switching}
\label{appendix:logic_switching}
\begin{figure}[!ht]
	\centerline{\includegraphics{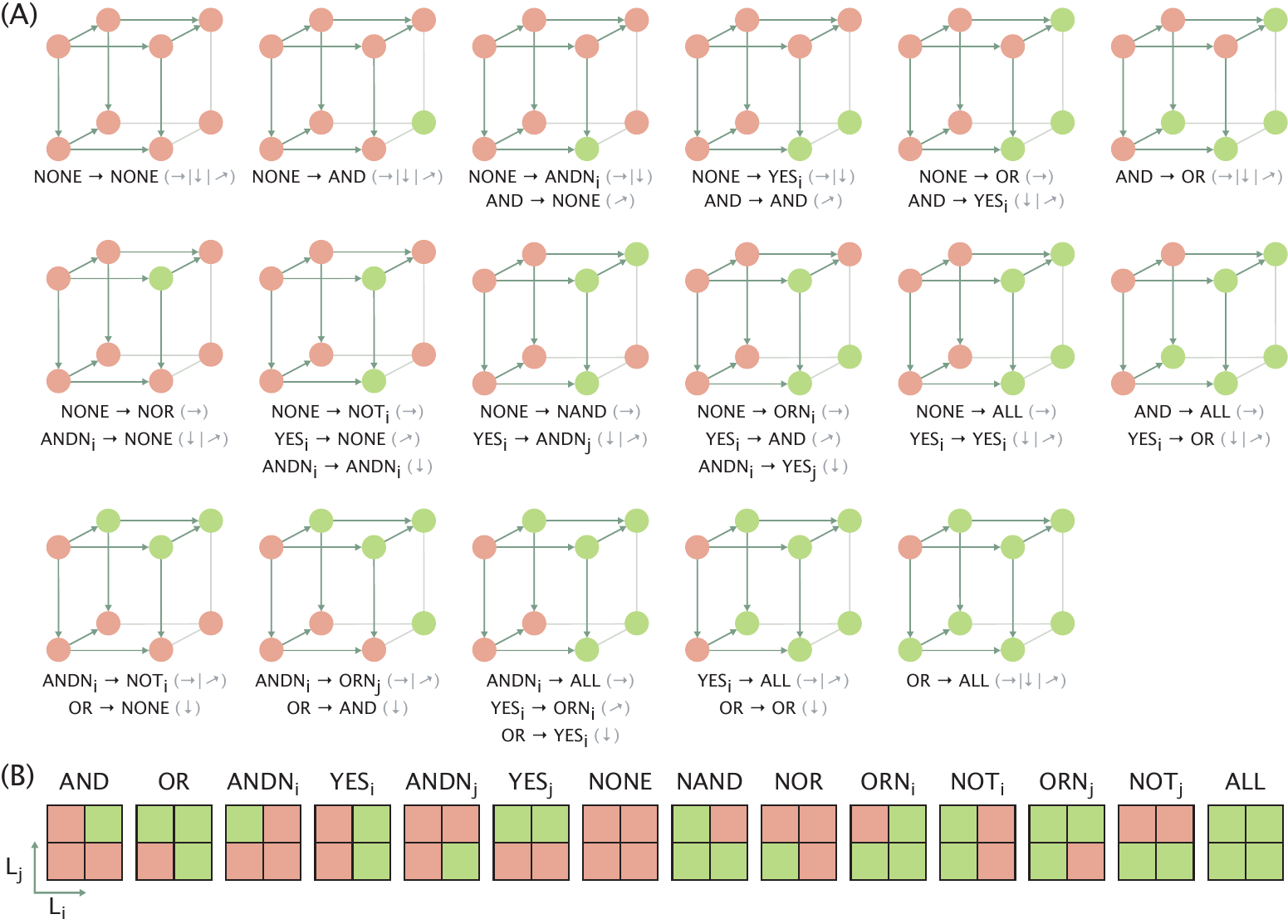}}
	\caption{\textbf{Functionally unique 3-ligand MWC gates and 
	possible schemes of logic switching.}
	(A) List of functionally unique 3-ligand MWC gates that have an
	inactive base state (in the absence of ligands). The set of logic switches that can be
	achieved by increasing the concentration of one of the ligands
	is listed on the bottom of each gate,
	with the gray arrows indicating the corresponding
	directions of increasing ligand concentration.
	Transitions with swapped
	labels ($\text{i} \leftrightarrow \text{j}$) are also possible and are not listed.
	Arrows corresponding to the ligand axes on different faces of the cube
	are included to assist the derivation of possible logic switches.
	(B) Schematics of 2-ligand gates adapted from 
	Figure \ref{fig:ThreeLigandPlusTransitions}D for convenience.
	}
	\label{figS:MWC_all}
\end{figure}

Here we describe how the table of all possible logic switches inducible by a third ligand
(Figure \ref{fig:ThreeLigandPlusTransitions}D) can be obtained from 
the list of MWC-compatible 3-ligand gates (Figure \ref{figS:MWC_all}), 
and also derive the parameter conditions
for AND$\to$OR and AND$\to$YES$_1$ logic switches.

As illustrated in Figure \ref{fig:ThreeLigandPlusTransitions}C, logic switching
can be achieved by increasing the concentration of any of the three ligands.
Following the same procedure, we 
iterate over the list of gates shown in Figure \ref{figS:MWC_all}A and
for each of them identify the set of possible logic switches.
The set of all logic switches present in Figure \ref{figS:MWC_all}A 
together constitute the entries of the table in Figure \ref{fig:ThreeLigandPlusTransitions}D.
Note that if a gate is compatible with the MWC framework, then
its logical complement is also compatible, and therefore,
the possibility of switching between two gates, $\text{Gate 1} \to \text{Gate 2}$,
implies the possibility of switching between their logical complements,
$\text{NOT (Gate 1)} \to \text{NOT (Gate 2)}$.

Now, we show how an MWC protein can exhibit the switching behaviors 
in Figure \ref{fig:exampleSwitches}B,D 
(AND$\to$OR and AND$\to$YES$_1$)
by saturating the concentration of the third ligand. We first consider
the behavior of the protein in the absence of the third ligand 
($[\text{L}_3] = 0$, with $\text{p}_{\text{active}}$ limits given in Figure
\ref{figS:MWC_constraints}A, left) and then consider how the protein acts at
the saturating concentration of the third ligand ($[\text{L}_3] \to \infty$, with
$\text{p}_{\text{active}}$ limits given in Figure \ref{figS:MWC_constraints}A, right).
With $[\text{L}_3] = 0$, the protein ignores the third ligand and behaves
identically to a protein with $\text{N}=2$ ligands. In the limit $[\text{L}_3] \to \infty$,
however, the protein behaves as if it only has two ligands with an altered
free energy difference $\Delta \text{\textepsilon}^\prime_{\text{AI}}$
between the active and inactive states given by
\begin{align} \label{eq:deltaEpsilonPrime}
\Delta \text{\textepsilon}^\prime_{\text{AI}} = \Delta \text{\textepsilon}_{\text{AI}} - \text{k}_\text{B} \text{T} \log \gamma_3.
\end{align}

Suppose that a protein acts as an AND gate when $[\text{L}_3] = 0$ and
transitions into an OR gate when $[\text{L}_3] \to \infty$, 
as in Figure \ref{fig:exampleSwitches}B. From Figure 
\ref{fig:MWC_simple}B, the MWC parameters must satisfy
\begin{align}
\label{eq:swap1}
\frac{1}{\gamma_1}, \frac{1}{\gamma_2} \ll \text{e}^{-\text{\textbeta} \Delta \text{\textepsilon}_{\text{AI}}} \ll \frac{1}{\gamma_1 \gamma_2}
\end{align}
in the absence of $\text{L}_3$ (AND behavior) and
\begin{align}
\label{eq:swap2}
1 \ll \text{e}^{-\text{\textbeta} \Delta \text{\textepsilon}^\prime_{\text{AI}}} \ll \frac{1}{\gamma_1}, \frac{1}{\gamma_2}
\end{align}
when $[\text{L}_3]$ is saturating (OR behavior). Using eqs~\ref{eq:deltaEpsilonPrime}, we can rewrite the condition \ref{eq:swap2} as
\begin{align}
\label{eq:swap2alt}
\frac{1}{\gamma_3}
\ll \text{e}^{-\text{\textbeta} \Delta \text{\textepsilon}_{\text{AI}}} \ll \frac{1}{\gamma_1 \gamma_3}, \frac{1}{\gamma_2 \gamma_3}.
\end{align}
Combining eq \ref{eq:swap1} and eq \ref{eq:swap2alt}, we find the second condition reported in Figure~\ref{fig:exampleSwitches}A, namely,
\begin{align} \label{eq:classSwitchingRequirement1}
\frac{1}{\gamma_1}, \frac{1}{\gamma_2}, \frac{1}{\gamma_3}
\ll \text{e}^{-\text{\textbeta} \Delta \text{\textepsilon}_{\text{AI}}} \ll \frac{1}{\gamma_1 \gamma_2}, \frac{1}{\gamma_1 \gamma_3}, \frac{1}{\gamma_2 \gamma_3}.
\end{align}
The first condition in Figure \ref{fig:exampleSwitches}A is then obtained by using the outer inequalities, that is,
\begin{align}
\frac{1}{\gamma_\text{k}} &\ll \frac{1}{\gamma_\text{i} \gamma_\text{j}} \Rightarrow \gamma_\text{i} \gamma_\text{j} \ll \gamma_\text{k} \quad \text{and} \\
\frac{1}{\gamma_\text{i}} &\ll \frac{1}{\gamma_\text{i} \gamma_\text{k}} \Rightarrow \gamma_\text{k} \ll 1.
\end{align}

Lastly, we derive the parameter conditions needed to achieve an AND$\to$YES$_1$ switching by saturating the third ligand. Conditions for the AND behavior in the absence of the third ligand are already known (eq \ref{eq:swap1}). To achieve a {YES$_1$} gate, $\text{p}_\text{active}$ at $[\text{L}_3] \to \infty$ needs to meet the following limits:
\begin{align}
\text{p}_{0,0, \infty} &= \frac{1}{1+ \gamma_3 \text{e}^{-\text{\textbeta} \Delta \text{\textepsilon}_{\text{AI}}}} \approx 0, \\
\text{p}_{0,\infty, \infty} &= \frac{1}{1+\gamma_2 \gamma_3 \text{e}^{-\text{\textbeta} \Delta \text{\textepsilon}_{\text{AI}}}} \approx 0, \\
\text{p}_{\infty,0, \infty} &= \frac{1}{1+ \gamma_1 \gamma_3 \text{e}^{-\text{\textbeta} \Delta \text{\textepsilon}_{\text{AI}}}} \approx 1, \\
\text{p}_{\infty,\infty, \infty} &= \frac{1}{1+ \gamma_1\gamma_2 \gamma_3 \text{e}^{-\text{\textbeta} \Delta \text{\textepsilon}_{\text{AI}}}} \approx 1.
\end{align}
These limits suggest constraints on $\Delta \text{\textepsilon}_{\text{AI}}$, which, combined with eq \ref{eq:swap1}, result in
\begin{align}
\label{eq:YESAND}
\frac{1}{\gamma_1}, \frac{1}{\gamma_2}, \frac{1}{\gamma_3}, \frac{1}{\gamma_2 \gamma_3} \ll  \text{e}^{-\text{\textbeta} \Delta \text{\textepsilon}_{\text{AI}}} \ll \frac{1}{\gamma_1 \gamma_2},\frac{1}{\gamma_1 \gamma_3}, \frac{1}{\gamma_2 \gamma_3},\frac{1}{\gamma_1 \gamma_2 \gamma_3}.
\end{align}
The outer inequalities, in turn, suggest conditions for the $\gamma$ parameters, namely,
\begin{align}
\frac{1}{\gamma_\text{i}} &\ll \frac{1}{\gamma_\text{i} \gamma_\text{k}} \Rightarrow \gamma_\text{k} \ll 1, \\
\frac{1}{\gamma_2 \gamma_3} &\ll \frac{1}{\gamma_1 \gamma_2} \Rightarrow \gamma_1 \ll \gamma_{2}, \\
\frac{1}{\gamma_2 \gamma_3} &\ll \frac{1}{\gamma_1 \gamma_3} \Rightarrow \gamma_1 \ll \gamma_{3}.
\end{align}
Accounting for these additional constraints, eq \ref{eq:YESAND} simplifies into
\begin{align}
\frac{1}{\gamma_1}, \frac{1}{\gamma_2 \gamma_3} \ll  \text{e}^{-\text{\textbeta} \Delta \text{\textepsilon}_{\text{AI}}} \ll \frac{1}{\gamma_1 \gamma_2},\frac{1}{\gamma_1 \gamma_3},
\end{align}
as shown in Figure \ref{fig:exampleSwitches}C.

\end{document}